\documentclass[%
  reprint,
  showpacs,preprintnumbers,
  amsmath,amssymb,
  aps,
  prb,
uplatex
]{revtex4-2}

\usepackage{graphicx}[]
\usepackage{dcolumn}
\usepackage{bm}
\usepackage{multirow}
\usepackage{braket}
\usepackage{color}
\usepackage{ulem}

\usepackage{hyperref}
\usepackage{url}

\begin{document}

\preprint{APS/123-QED}

\title{
Helicity locking of square skyrmion crystal in a centrosymmetric lattice system\\ without vertical mirror symmetry
}

\author{Satoru Hayami and Ryota Yambe}
\affiliation{
Department of Applied Physics, The University of Tokyo, Tokyo 113-8656, Japan
}
 
\begin{abstract}
We theoretically investigate the stability of a square skyrmion crystal (SkX) in a centrosymmetric tetragonal lattice structure with the emphasis on the role of the magnetic anisotropy arising from the absence of vertical mirror symmetry. 
Our analysis is based on an effective bilinear and biquadratic model in momentum space, which is a canonical model for itinerant magnets in a weak-coupling regime. 
By performing the simulated annealing for the model on the two-dimensional square lattice, we find that the off-diagonal spin component in the interaction, which becomes nonzero when the vertical mirror symmetry is broken, gives rise to the square SkX with a definite helicity in an external magnetic field. 
We show that the helicity of the centrosymmetric SkXs is determined by the competition between the off-diagonal and diagonal anisotropic interactions, the latter of which appears in the discrete fourfold-rotational lattice structure. 
Furthermore, we discuss helicity-dependent physical phenomena by introducing odd-parity multipoles, where electric (magnetic) and electric (magnetic) toroidal multipoles are sources of an antisymmetric spin polarization and an Edelstein effect (a magnetoelectric effect). 
We also discuss the stability of the SkXs with different helicities in a magnetic field rotation.  
Our results provide a way of engineering the helicity-locked SkXs by the symmetric anisotropic interaction in centrosymmetric magnets, which is distinct from that by the antisymmetric Dzyaloshinskii-Moriya interaction in noncentrosymmetric magnets. 
\end{abstract}
\maketitle

\section{Introduction}

Noncollinear and noncoplanar spin configurations have been extensively studied in condensed matter physics in recent decades, as they exhibit unique physical phenomena and potential device applications~\cite{Nagaosa_RevModPhys.82.1539,Xiao_RevModPhys.82.1959,nagaosa2013topological,Baltz_RevModPhys.90.015005,smejkal2021anomalous}. 
Among them, a noncoplanar spin configuration with a nonzero topological (skyrmion) number, which is known as a magnetic skyrmion, is one of the central subjects~\cite{Bogdanov89,Bogdanov94,rossler2006spontaneous,Muhlbauer_2009skyrmion,yu2010real,yu2011near,seki2012observation,nagaosa2013topological}. 
Owing to its topologically protected structure, the magnetic skyrmion behaves as a particle that can be manipulated, which leads to information carries in next generation spintronic devices~\cite{pfleiderer2010single,schulz2012emergent,iwasaki2013current,fert2013skyrmions,zhang2015magnetic,zhou2015dynamically,zhang2020skyrmion,PhysRevLett.127.067201}. 
Moreover, a periodic alignment of the magnetic skyrmion, i.e., a magnetic skyrmion crystal (SkX), has also drawn considerable attention, since it shows various unconventional macroscopic response properties related to topological and vortex spin textures in both metals and insulators, such as the topological Hall and Nernst effects
~\cite{Neubauer_PhysRevLett.102.186602,Hamamoto_PhysRevB.92.115417,Gobel_PhysRevB.95.094413,Saha_PhysRevB.60.12162,kurumaji2019skyrmion,Hirschberger_PhysRevLett.125.076602,Shiomi_PhysRevB.88.064409}, the magnetoelectric effect~\cite{seki2012observation,white2012electric,okamura2013microwave,Mochizuki_PhysRevB.87.134403,tokura2014multiferroics,mochizuki2015dynamical,Christensen_PhysRevX.8.041022,Gobel_PhysRevB.99.060406}, and the nonreciprocal transport~\cite{Seki_PhysRevB.93.235131,giordano2016spin,tokura2018nonreciprocal,yokouchi2018current,Hoshino_PhysRevB.97.024413,seki2020propagation,hayami2020phase}. 

The SkXs have been ubiquitously found in materials with a variety of lattice structures including both noncentrosymmetric and centrosymmetric structures~\cite{Tokura_doi:10.1021/acs.chemrev.0c00297}. 
The appearance of the SkXs in the noncentrosymmetric lattice structures is mainly owing to the Dzyaloshinskii-Moriya (DM) interaction denoted as $\bm{D} \cdot (\bm{S}_i \times \bm{S}_j)$ between two spins [$\bm{D}$ is so-called the DM vector and $\bm{S}_{i(j)}$ is the spin at site $i(j)$], which arises from the relativistic spin-orbit coupling without the inversion symmetry at the bond center~\cite{dzyaloshinsky1958thermodynamic,moriya1960anisotropic,rossler2006spontaneous}. 
Although the form of the DM vector depends on the point group symmetry, such as the mirror and rotational symmetries in addition to the spatial inversion symmetry in crystals, it is recognized that the DM interaction plays an important role in the stabilization of the SkXs in chiral~\cite{Muhlbauer_2009skyrmion,yu2010real,yu2011near,seki2012observation,Adams2012,Seki_PhysRevB.85.220406,tokunaga2015new,karube2016robust,Li_PhysRevB.93.060409,kakihana2018giant,kakihana2019unique,hayami2021field}, polar~\cite{heinze2011spontaneous,kezsmarki_neel-type_2015,Kurumaji_PhysRevLett.119.237201}, and other noncentrosymmetric magnets~\cite{nayak2017discovery,peng2020controlled}. 

In the centrosymmetric lattice structures, the SkXs are stabilized by considering the frustrated exchange interaction~\cite{Okubo_PhysRevLett.108.017206,leonov2015multiply,Lin_PhysRevB.93.064430,Hayami_PhysRevB.93.184413,batista2016frustration,Lin_PhysRevLett.120.077202,Hayami_PhysRevB.103.224418,hayami2022skyrmion} and effective long-range interactions originating from the spin-charge coupling in itinerant electron systems~\cite{hayami2021topological}, such as the Ruderman-Kittel-Kasuya-Yosida interaction~\cite{Ruderman,Kasuya,Yosida1957,Wang_PhysRevLett.124.207201,mitsumoto2021replica,mitsumoto2021skyrmion} and the multiple-spin interactions~\cite{Ozawa_PhysRevLett.118.147205,Hayami_PhysRevB.95.224424,Hayami_PhysRevB.99.094420,hayami2020multiple,Eto_PhysRevB.104.104425,Hayami_10.1088/1367-2630/ac3683} with and without single-ion anisotropy.
In contrast to the DM-interaction mechanism, the spin interactions in these mechanisms are characterized by the isotropic ones at least for the two spin components, which are exemplified by $S^x_{i}S^x_{j}+S^y_{i}S^y_{j}$
, $\bm{S}_{i}\cdot \bm{S}_{j}$, $\bm{S}_{\bm{q}}\cdot \bm{S}_{-\bm{q}}$, and $(\bm{S}_{\bm{q}}\cdot \bm{S}_{-\bm{q}})^2$, where $\bm{S}_{\bm{q}}$ is the Fourier component of $\bm{S}_i$ with wave vector $\bm{q}$.  
Such a stabilization mechanism based on the isotropic exchange interactions represents a universal feature irrespective of the details of the lattice symmetry. 
Moreover, the isotropic exchange interactions result in the degeneracy of the SkXs and anti SkXs with a different sign of the skyrmion number, which provides a possibility of new types of the topological spin orderings~\cite{Okubo_PhysRevLett.108.017206,mitsumoto2021skyrmion}. 

Meanwhile, magnetic anisotropy that arises from the rotational and mirror symmetry breakings in the specific lattice structures while keeping the inversion symmetry also becomes a source of the SkXs~\cite{yambe2022effective}. 
Although the form of the spin interactions depends on a way of breakings of the lattice symmetry, such an effect can appear in any discrete lattice systems via the spin-orbit coupling. 
For instance, the bond-dependent anisotropic exchange interaction owing to the discrete rotational symmetry gives rise to the square-shaped SkXs in the tetragonal system~\cite{Hayami_PhysRevLett.121.137202,Hayami_PhysRevB.103.024439,Wang_PhysRevB.103.104408,hayami2022multiple} and the triangular-shaped SkXs in the hexagonal system~\cite{amoroso2020spontaneous,Hayami_PhysRevB.103.054422,amoroso2021tuning}. 
This bond-dependent interaction often plays a similar role to the dipole-dipole interaction in the stabilization of the SkXs~\cite{Utesov_PhysRevB.103.064414,utesov2021mean}.
Besides, bond-dependent anisotropic exchange interaction originating from the breaking of the horizontal mirror symmetry also leads to the SkXs in the trigonal system~\cite{amoroso2020spontaneous,yambe2021skyrmion,amoroso2021tuning}. 
More recently, it was shown that the staggered DM interaction that originates form the breaking of the local inversion symmetry becomes the origin of the SkX in centrosymmetric magnets~\cite{Hayami_PhysRevB.105.014408,lin2021skyrmion}.
These mechanisms on the basis of the anisotropic interactions account for an anisotropic directional response of the SkXs in centrosymmetric magnets against an external magnetic field~\cite{Hirschberger_10.1088/1367-2630/abdef9}. 

In addition, magnetic anisotropy plays an important role in lifting the degeneracy between the helicity and the vorticity of the centorsymmetric SkXs. 
For example, the bond-dependent anisotropic interaction in the hexagonal systems lifts the degeneracy between the N\'eel, Bloch, and anti SkXs~\cite{amoroso2020spontaneous,Hayami_PhysRevB.103.054422,amoroso2021tuning}. 
Similarly, their degeneracy is lifted in the tetragonal~\cite{Hayami_doi:10.7566/JPSJ.89.103702,Hayami_PhysRevLett.121.137202,Hayami_PhysRevB.103.024439,Wang_PhysRevB.103.104408} and trigonal systems
~\cite{amoroso2020spontaneous,yambe2021skyrmion,amoroso2021tuning}. 
In fact, the SkXs observed in centrosymmetric magnets have definite helicity and vorticity, which indicates the importance of the anisotropic interaction and dipole-dipole 
interaction~\cite{zhang2017skyrmion,kurumaji2019skyrmion,hirschberger2019skyrmion,khanh2020nanometric}. 
As the magnetic anisotropy is different under the different lattice symmetry, it is desired to examine important anisotropic interactions to stabilize the SkXs in each lattice symmetry, which will give a guideline to search for further SkX-hosting materials based on the crystallographic point groups. 

In the present study, we aim at exploring the SkXs induced by the magnetic anisotropy in a discrete lattice system. 
We focus on the effect of the magnetic anisotropy that arises from the lacking of vertical mirror symmetry (vertical twofold rotational symmetry) on the stabilization of the square SkX under the centrosymmetric tetragonal $C_{4{\rm h}}$ point group system. 
Specifically, we examine an effective spin model of itinerant magnets consisting of itinerant electrons and localized spins, which includes two types of bond-dependent anisotropic interactions satisfying the $C_{4{\rm h}}$ point group symmetry: 
One is the spin-diagonal anisotropy in the presence of the fourfold rotational symmetry and the other is the spin-off-diagonal anisotropy in the absence of the vertical mirror symmetry. 
By numerically analyzing the effective spin model at low temperatures while changing the biquadratic interaction, anisotropic interaction, and magnetic field, we find that the interplay between the magnetic anisotropic and biquadratic interactions induces the square SkXs in an external magnetic field. 
We construct the low-temperature phase diagrams while changing the biquadratic interaction, anisotropic interaction, and magnetic field to demonstrate a stabilization tendency of not only the SkX but also other double-$Q$ states in a systematic manner. 
We show that the helicity of the SkXs is locked depending on the ratio and sign of two magnetic anisotropic interactions. 
According to the different helicity, the SkXs accompany different types of odd-parity multipoles, which are related to the emergence of a linear magnetoelectric effect, an antisymmetric spin polarization, and an Edelstein effect~\cite{edelstein1990spin}. 
We also discuss the stability of the SkXs with the different helicities when the magnetic field is rotated from the out-of-plane to inplane directions. 
Our result provides a close relation between the helicity of the SkXs and the magnetic anisotropic interactions in centrosymmetric itinerant magnets, which will be applicable to the other point groups without the vertical mirror symmetry, such as $C_4$, $S_{\rm 4}$, $C_{\rm 6}$, $S_{\rm 6}$, and $C_{\rm 6h}$.

The remainder of this paper is structured as follows. 
In Sec.~\ref{sec:Effective spin model}, we present an effective spin model incorporating the effect of the vertical mirror symmetry breaking in the centrosymmetric tetragonal system. 
We also outline numerical simulations based on the simulated annealing. 
In Sec.~\ref{sec:Result}, we discuss the low-temperature phase diagrams in the effective spin model, a helicity locking in the SkX, a relation to odd-parity multipoles, and an effect of the magnetic field rotation. 
Finally, Sec.~\ref{sec:Summary} concludes this article. 

\section{Effective spin model}
\label{sec:Effective spin model}

To investigate the effect of the vertical mirror symmetry breaking on the SkXs, we consider the tetragonal system under the crystallographic point group $C_{\rm 4h}$. 
Specifically, we analyze a phenomenological spin model with the anisotropic interactions defined in momentum space on the two-dimensional square lattice in the $xy$ plane, whose Hamiltonian is given by 
\begin{align}
\label{eq:Model}
\mathcal{H}=  &-2J\sum_{\eta=1,2}
\sum_{\mu,\nu}\Gamma^{\mu\nu}_{\bm{Q}_{\eta}} S^\mu_{\bm{Q}_{\eta}} S^\nu_{-\bm{Q}_{\eta}} \nonumber \\ 
&+\frac{2K}{N} \sum_{\eta=1,2}  \Bigg(\sum_{\mu,\nu}\Gamma^{\mu\nu}_{\bm{Q}_{\eta}} S^\mu_{\bm{Q}_{\eta}} S^\nu_{-\bm{Q}_{\eta}}\Bigg)^2  
- \sum_i \bm{H} \cdot \bm{S}_i,
\end{align}
where 
\begin{align}
\label{eq:Gamma1}
\Gamma_{\bm{Q}_1}&=\left(
\begin{array}{ccc}
I-I^{v} & I^{xy} & 0\\
I^{xy} & I+I^{v} &0 \\
0 & 0 & I^{z}
\end{array}
\right), \\
\label{eq:Gamma2}
\Gamma_{\bm{Q}_2}&=\left(
\begin{array}{ccc}
I+I^{v} & -I^{xy} & 0\\
-I^{xy} & I-I^{v} &0 \\
0 & 0 & I^{z}
\end{array}
\right).   
\end{align}
The Hamiltonian in Eq.~(\ref{eq:Model}) consists of three terms. 
The first term represents the bilinear exchange interaction with the coupling constant $J$ and the second term represents the biquadratic exchange interaction with the coupling constant $K$ for the wave vectors $\bm{Q}_1=(\pi/3,0)$ and $\bm{Q}_2=(0,\pi/3)$ (the lattice constant is taken as unity); $\bm{S}_{\bm{Q}_{\eta}}$ is the $\bm{Q}_{\eta}$ component of the spins obtained by the Fourier transform of the classical localized spin $\bm{S}_i$ with $|\bm{S}_i|=1$ and $N$ is the system size. 
It is noted that $\bm{Q}_1$ and $\bm{Q}_2$ are connected by the fourfold rotational symmetry of the square-lattice geometry. 
The interaction tensors in the $\bm{Q}_1$ and $\bm{Q}_2$ channels have an anisotropic form in terms of the spin components $\mu,\nu=x,y,z$, as given in Eqs.~(\ref{eq:Gamma1}) and (\ref{eq:Gamma2}), which are obtained by the symmetry argument for the point group $C_{\rm 4h}$; 
there are four independent model parameters, $I$, $I^{v}$, $I^{xy}$, and $I^z$, and $\Gamma^{\mu\nu}_{\bm{Q}_\eta}$ is invariant under the 
twofold rotational, the space inversion, and the horizontal mirror symmetries, where $\Gamma_{\bm{Q}_1}$ and $\Gamma_{\bm{Q}_2}$ are connected by the fourfold rotation. 
Among them, the anisotropic interactions $I^z \neq I$ and $I^v$ become nonzero even in the tetragonal point group with the vertical mirror symmetry, such as $D_{\rm 4h}$. 
Meanwhile, the other anisotropic interaction $I^{xy}$ only appears when both the vertical mirror and vertical twofold rotational symmetries are broken. 
The third term represents the Zeeman coupling to an external magnetic field in the form of $\bm{H}=H(\sin \theta \cos \phi, \sin\theta \sin \phi, \cos \theta)$. 

The effective bilinear and biquadratic model in momentum space in Eq.~(\ref{eq:Model}) is one of the canonical model to discuss the multiple-$Q$ instabilities in itinerant magnets~\cite{Hayami_PhysRevB.95.224424,hayami2020multiple,Hayami_PhysRevB.103.024439,Hayami_PhysRevB.103.054422,hayami2021topological}. 
The bilinear and biquadratic interactions are derived from the classical Kondo lattice model consisting of the itinerant electrons and localized spins by tracing out the itinerant electron degree of freedom. 
According to the perturbative expansion in terms of the exchange coupling between the itinerant electron and localized spins, the bilinear and biquadratic interactions are proportional to the second and fourth orders of the exchange coupling, respectively. 
The choice of the interactions at the wave vectors $\bm{Q}_1$ and $\bm{Q}_2$ is based on the assumption that the magnetic susceptibility of itinerant electrons shows maximum peaks at the corresponding wave vectors in the presence of the nested Fermi surfaces by $\bm{Q}_1$ and $\bm{Q}_2$~\cite{Akagi_PhysRevLett.108.096401,Hayami_PhysRevB.90.060402,Ozawa_doi:10.7566/JPSJ.85.103703,Hayami_PhysRevB.95.224424}. 
We neglect the contributions from the other multi-spin interactions for the same reason. 

The multiple-$Q$ instabilities in the effective model in Eq.~(\ref{eq:Model}) have been discussed for the case of the centrosymmetric tetragonal point group $D_{\rm 4h}$, i.e., $I^{xy}=0$, in the previous literatures~\cite{Hayami_PhysRevB.95.224424,Hayami_doi:10.7566/JPSJ.89.103702,Yasui2020imaging,Su_PhysRevResearch.2.013160,Hayami_PhysRevB.103.024439,seo2021spin}. 
In the case of the isotropic spin interaction $I=I^{z}$ and $I^v=0$, only topologically-trivial single-$Q$ and
double-$Q$ states appear while changing $K$ and $H$ in the unit of $J$~\cite{Hayami_PhysRevB.95.224424}. 
Subsequently, it was shown that the square SkX is realized by incorporating the effect of $I^{v}$ in addition to $I^z>I$, $K$, and $H$ for $\theta=0$~\cite{Yasui2020imaging,Hayami_PhysRevB.103.024439}; the Bloch (N\'eel) SkX is stabilized for $I^v>0$ ($I^v<0$), although there is a degeneracy between the SkXs and anti SkXs.

In the following, we discuss the role of the anisotropic interaction $I^{xy}$ characteristics of the point group $C_{\rm 4h}$, which appears when the vertical mirror symmetry of the point group $D_{\rm 4h}$ is lost while keeping the inversion symmetry. 
In order to investigate the low-temperature phase diagram of the model including $I^{xy}$ in Eq.~(\ref{eq:Model}), we carry out the simulated annealing by means of Monte Carlo simulations. 
The simulations have been done by following the manner in Ref.~\onlinecite{Hayami_PhysRevB.103.024439} with the same final temperature as $T=0.01$. 
In the simulation processes, we reduce the temperature with a rate of $\alpha=0.99995-0.99999$ at the $n$th Monte Carlo step from a random spin configuration at a high temperature $T_0=1-10$.
At the final temperature, we perform $10^5$-$10^6$ Monte Carlo sweeps for measurements. 
We also start the simulations from the spin configurations obtained at low temperatures when determining the phase boundaries in the phase diagram. 
We set $J=1$ as the energy unit of the model and $I=1$ as the unit of the anisotropic form factor. 
We fix $I^z =1.2$, as the situation satisfying $I^z>I$ tends to stabilize the SkX. 
We change the other parameters $K$, $I^v$, and $I^{xy}$ in the interaction tensors and the magnetic field $\bm{H}$ to discuss the stability of the square SkX systematically. 
The system size is taken for $N=96^2$ spins. 
It is noteworthy that the effect of thermal fluctuations on the model with long-range interactions similar to that in Eq.~(\ref{eq:Model}) has recently been investigated, where the multiple-$Q$ states stabilized at low temperatures tend to survive at finite temperatures~\cite{kato2022magnetic}.

We identify each magnetic phase in the following results by examining the spins in momentum space. 
For that purpose, we calculate the spin structure factor for the $\mu=x,y,z$ component defined by 
\begin{align}
S_s^{\mu}(\bm{q})&= \frac{1}{N} \sum_{j,l} S_j^{\mu} S_l^{\mu}  e^{i \bm{q}\cdot (\bm{r}_j-\bm{r}_l)}, 
\end{align}
where $\bm{r}_j$ is the position vector at site $j$. 
We also use the notation $S_s^{xy}(\bm{q})= S_s^{x}(\bm{q})+S_s^{y}(\bm{q})$. 
Then, the magnetic moments at the $\bm{Q}_\eta$ component are expressed as $m^{\mu}_{\bm{Q}_\eta}=\sqrt{S^{\mu}_s(\bm{Q}_\eta)/
N}$. 
The net magnetization is defined by $M_{\mu}=(1/N)\sum_i S_i^{\mu}$. 

In addition, we evaluate the scalar chirality $\chi_0$ to investigate whether the obtained spin configurations are topologically nontrivial, which is calculated from 
\begin{align}
\chi_0&= \frac{1}{N}\sum_{i} \chi_i, \\
\chi_i&=\sum_{\delta=\pm1}\bm{S}_i \cdot (\bm{S}_{i+\delta \hat{x}}\times \bm{S}_{i+\delta \hat{y}}),  
\end{align}
where $\hat{x}$ ($\hat{y}$) is the unit vector in the $x$ ($y$) direction~\cite{Yi_PhysRevB.80.054416}. 
This quantity becomes nonzero for noncoplanar spin configurations, which is related to a quantized skyrmion number. 
Complementaliry, we also compute the skyrmion number defined by~\cite{BERG1981412}
\begin{align}
\label{eq:nsk_num}
n_{\rm sk}=\frac{1}{2\pi N_m}\sum_{i,\delta=\pm 1}  \tan^{-1} \frac{\bm{S}_i \cdot (\bm{S}_j \times \bm{S}_k)}{1+\bm{S}_i \cdot \bm{S}_j+\bm{S}_j \cdot \bm{S}_k+\bm{S}_k \cdot \bm{S}_i},
\end{align} 
where $N_m$ is the number of magnetic unit cell in the lattice system, and $j=i+\delta \hat{x}$ and $k=i+\delta \hat{y}$; the range of the arctangent is set as $[-\pi, \pi)$. 
$n_{\rm sk}$ is quantized at $-1$ ($+1$) for the SkX (anti SkX).

\section{Result}
\label{sec:Result}

In this section, we discuss the results obtained by the simulated annealing for the effective spin model in Eq.~(\ref{eq:Model}) on the square lattice. 
We first present a magnetic phase diagram with $I^{xy}$ but without $I^v$ while changing the biquadratic interaction $K$ and the out-of-plane field $H$ ($\theta=0$) in Sec.~\ref{sec:Skyrmion crystal in an out-of-plane field}.  
We find that the square SkX is induced when $K$ and $H$ are nonzero. 
We also discuss the stability of the square SkX while changing $I^{xy}$ for fixed $K$. 
In Sec.~\ref{sec:Helicity locking of skyrmion crystal}, we discuss the change of the spin configurations in the presence of both $I^{xy}$ and $I^v$. 
We show that the helicity of the SkX is locked depending on the ratio and sign of $I^{xy}$ and $I^v$. 
Then, in Sec.~\ref{sec:Relation to odd-parity multipoles}, we show that the different helicity makes different types of odd-parity multipoles active, which is related to the linear magnetoelectric effect and the antisymmetric spin polarization. 
Lastly, we discuss the stability of the SkX when the magnetic field is tilted from the out-of-plane to the inplane directions in Sec.~\ref{sec:Stability of skyrmion crystal in a rotated field}. 
It is noted that the following results do not change in the three-dimensional 
layered square lattice structure with the two-dimensional ordering vectors.

\subsection{Skyrmion crystal in an out-of-plane field}
\label{sec:Skyrmion crystal in an out-of-plane field}

\begin{figure}[t!]
\begin{center}
\includegraphics[width=1.0 \hsize ]{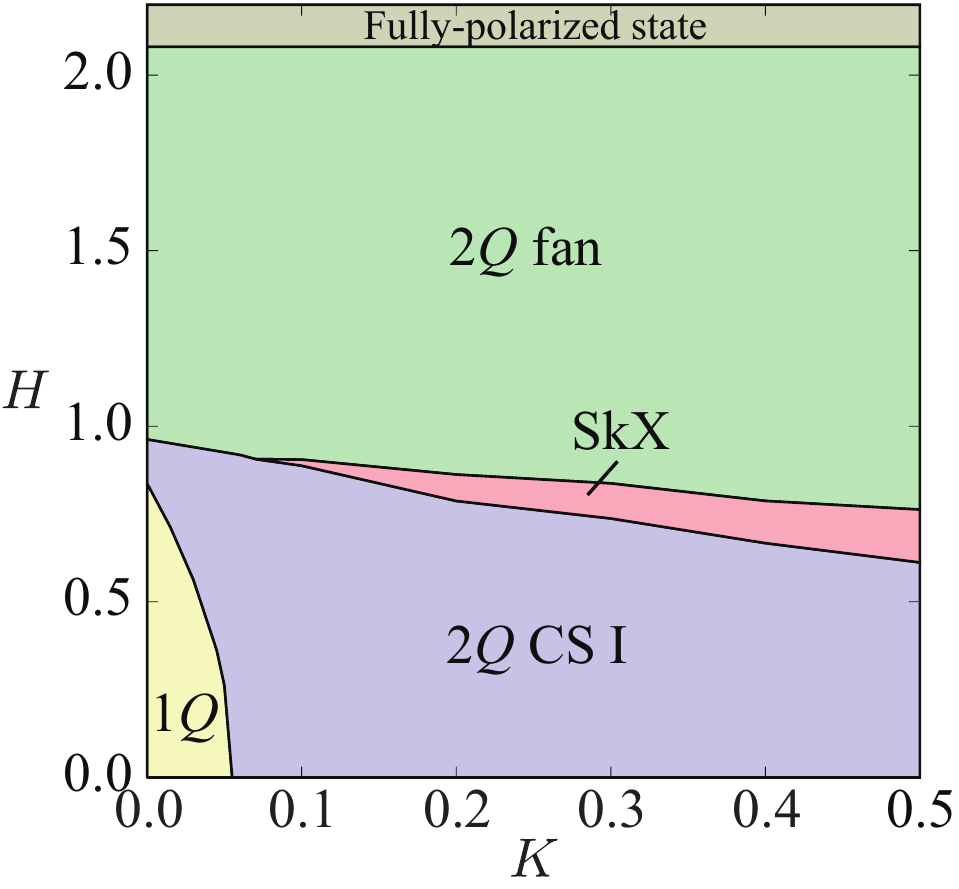} 
\caption{
\label{fig:PD}
Phase diagram while changing the biquadratic interaction $K$ and the out-of-plane magnetic field $H$ ($\theta=0$) at $I^z=1.2$, $I^{xy}=0.05$, and $I^v=0$ obtained by the simulated annealing. 
1$Q$ (2$Q$) represents the single-$Q$ (double-$Q$) state.
CS represents the chiral stripe state. 
}
\end{center}
\end{figure}

\begin{figure*}[t!]
\begin{center}
\includegraphics[width=1.0 \hsize ]{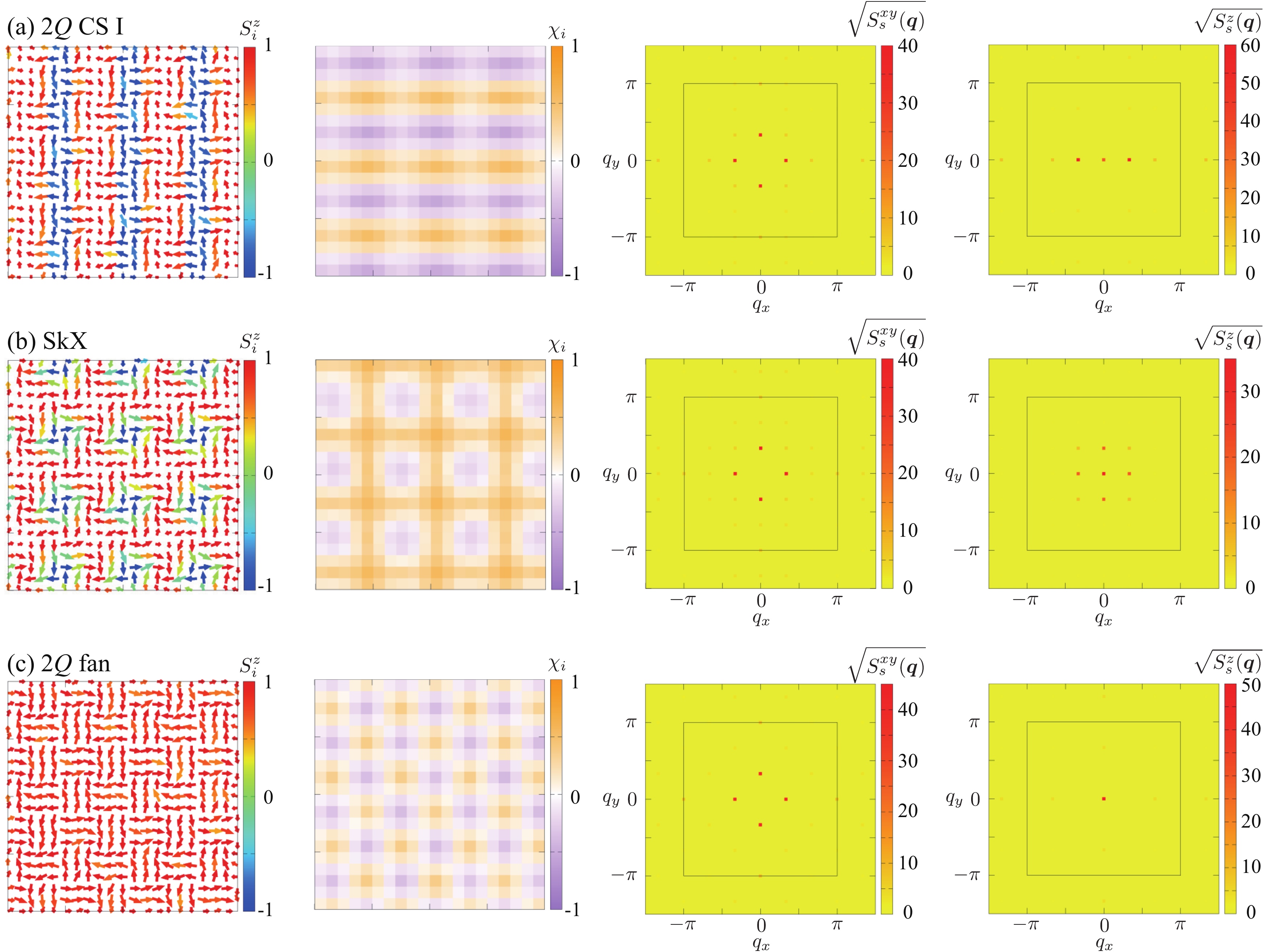} 
\caption{
\label{fig:spin}
Left: Snapshots of the spin configurations in (a) the 2$Q$ chiral stripe (CS) I state for $H=0.7$, (b) the SkX for $H=0.8$, and (c) the 2$Q$ fan state for $H=0.9$ at $K=0.2$, $I^z=1.2$, $I^{xy}=0.05$, and $I^v=0$. 
The direction and the color of the arrows represent the $xy$ and $z$ components of the spin moment, respectively. 
Middle left: the scalar chirality $\chi_i$ calculated from the left panel. 
Middle right and right: The square root of the $xy$ and $z$ components of the spin structure factor. 
Black squares represent the first Brillouin zone.
}
\end{center}
\end{figure*}

\begin{figure}[t!]
\begin{center}
\includegraphics[width=1.0 \hsize ]{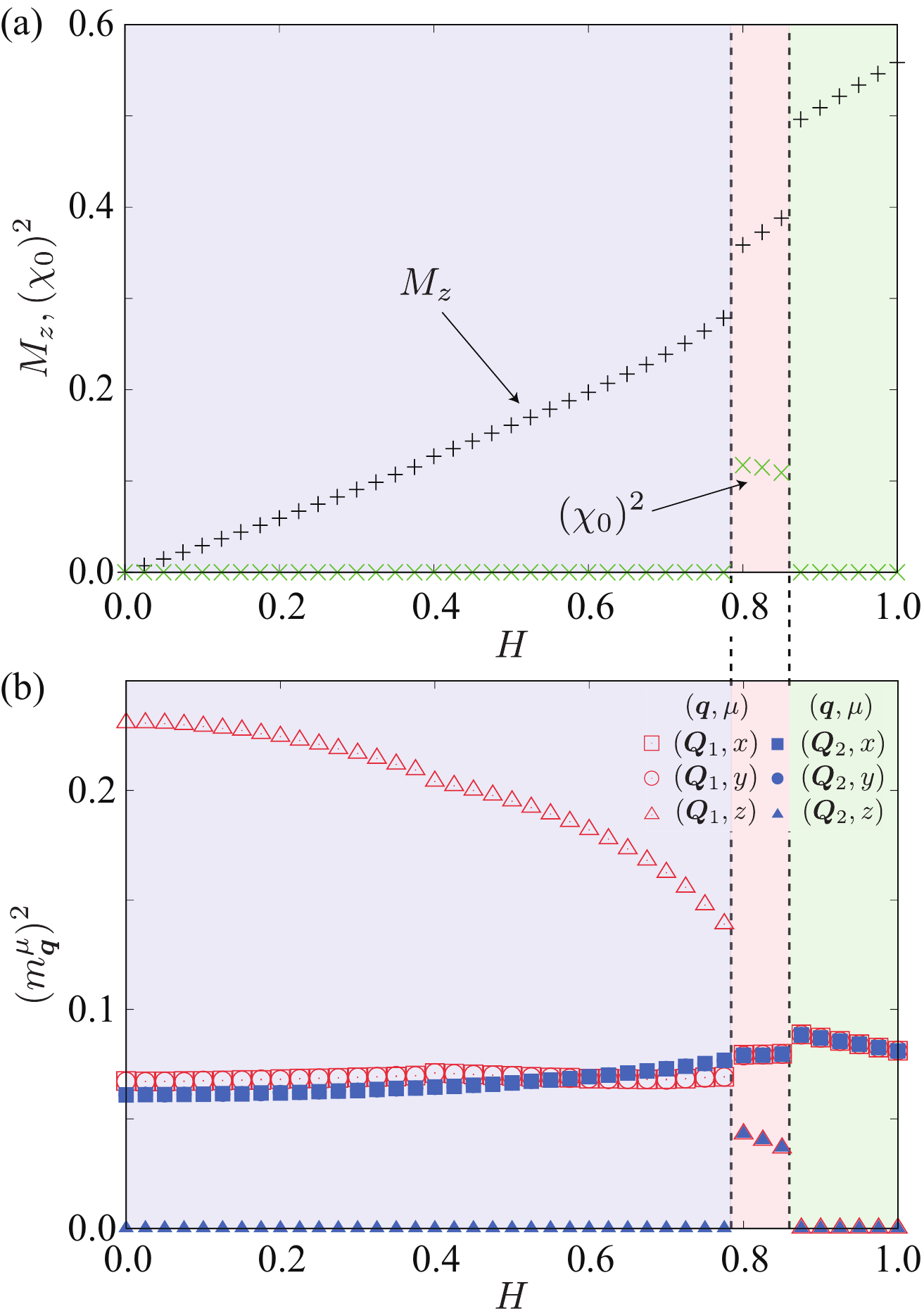} 
\caption{
\label{fig:Mq}
$H$ dependences of (a) the magnetization $M_z$, the scalar chirality $(\chi_0)^2$, and (b) the $\bm{Q}_1$ and $\bm{Q}_2$ components of the magnetic moments, $m^\mu_{\bm{Q}_1}$ and $m^\mu_{\bm{Q}_2}$, for $\mu=x,y,z$ at $K=0.2$, $I^z=1.2$, $I^{xy}=0.05$, and $I^v=0$. 
}
\end{center}
\end{figure}

We discuss the low-temperature phase diagram of the effective spin model in Eq.~(\ref{eq:Model}) by performing the simulated annealing. 
We consider nonzero $I^{xy}$ but $I^v=0$ in this section, although it has been already known that nonzero $I^v$ without $I^{xy}$ can stabilize the Bloch or N\'eel SkX, where the stability region of the SkX while changing the model parameters is presented only for a few set of model parameters~\cite{Yasui2020imaging,Hayami_PhysRevB.103.024439}. 
(It is noted that the model for $I^{xy} \neq 0$ and $I^v=0$ is transformed to that for $I^{xy}=0$ and $I^v \neq 0$ by permutating the inplane spins appropriately.) 
We also take the magnetic field along the $z$ direction $\bm{H}=(0,0,H)$ by taking $\theta=0$, since the inplane field under the easy-axis anisotropic interaction ($I^z>I$) tends to destabilize the SkX~\cite{hayami2020multiple,Hayami_PhysRevB.103.024439}.

Figure~\ref{fig:PD} shows the phase diagram in the $K$-$H$ plane at $T=0.01$, $I^z=1.2$, $I^{xy}=0.05$, and $I^v=0$. 
There are four phases in addition to the fully-polarized state along the field direction for $H \gtrsim 2.1$. 
The spin and chirality configurations in three double-$Q$ phases are shown in Fig.~\ref{fig:spin}.  
In the low-field region, the single-$Q$ state appears for small $K$, whose spin configuration is characterized by the elliptical spiral one; 
the spiral plane lies on the [110] or [$1\bar{1}$0] plane. 
While increasing $K$, the second-$Q$ component in the spin structure factor is induced and developed, and then, the double-$Q$ chiral stripe (CS) I state is stabilized. 
The $xy$ component of the spin structure factor shows the double-$Q$ peaks at $\bm{Q}_1$ and $\bm{Q}_2$ with different intensities, while its $z$ component exhibits the single-$Q$ peak at $\bm{Q}_1$, as shown in the right two panels in Fig.~\ref{fig:spin}(a). 
Then, the spin configuration obtained by the simulated annealing as shown in the left panel of Fig.~\ref{fig:spin}(a) is well described by
\begin{align}
\label{eq:2QCS}
\bm{S}_i \propto 
\left(
    \begin{array}{c}
    \cos \bm{Q}_1 \cdot \bm{r}_i+ b \cos \bm{Q}_2 \cdot \bm{r}_i \\
    \cos \bm{Q}_1 \cdot \bm{r}_i- b \cos \bm{Q}_2 \cdot \bm{r}_i \\
    a_z \sin \bm{Q}_1\cdot \bm{r}_i + \tilde{M}_z
          \end{array}
  \right)^{\rm T},
\end{align}
where $a_z$, $b$, and $\tilde{M}_z$ are parameters depending on the model parameters ($\tilde{M}_z=0$ when $H=0$). 
T in Eq.~(\ref{eq:2QCS}) represents the transpose of the vector. 
Thus, the double-$Q$ CS I state is represented by the superposition of the elliptical spiral wave along the $\bm{Q}_1$ direction and the sinusoidal wave along the $\bm{Q}_2$ direction. 
Owing to the presence of $I^{xy}$, the spiral plane and the sinusoidal oscillating direction lie on the [110] or $[1\bar{1}0]$ plane. 
Simultaneously, the double-$Q$ CS I state accompanies the scalar chirality density wave along the $\bm{Q}_2$ direction without a net component as shown in the middle left panel of Fig.~\ref{fig:spin}(a), which reflects the noncoplanar spin textures in Eq.~(\ref{eq:2QCS}). 
This double-$Q$ CS I state has been discussed in itinerant electron models on various lattices, such as the square~\cite{Ozawa_doi:10.7566/JPSJ.85.103703,Hayami_PhysRevB.95.224424,yambe2020double,Hayami_PhysRevB.103.024439}, triangular~\cite{Hayami_PhysRevB.95.224424,hayami2020multiple,Hayami_PhysRevB.103.054422}, and cubic~\cite{Okumura_PhysRevB.101.144416} lattices. 

While increasing $H$, the SkX appears for $K \gtrsim 0.07$, as shown in the phase diagram in Fig.~\ref{fig:PD}. 
The real-space spin configuration is characterized by the periodic array of the skyrmion in a square-lattice way, as shown in the left panel of Fig.~\ref{fig:spin}(b). 
This indicates the emergence of the square SkX~\cite{Hayami_PhysRevB.103.024439,Utesov_PhysRevB.103.064414,Wang_PhysRevB.103.104408}. 
The spin structure factor exhibits the double-$Q$ peaks with equal intensity at $\bm{Q}_1$ and $\bm{Q}_2$ in both $xy$ and $z$ components, as shown in the right two panels of Fig.~\ref{fig:spin}(b). 
The spin ansatz of the square SkX is given by the superposition of two elliptical spiral waves along the $\bm{Q}_1$ and $\bm{Q}_2$ directions as
\begin{align}
\label{eq:2QSkX}
\bm{S}_i \propto 
\left(
    \begin{array}{c}
    \cos \bm{Q}_1 \cdot \bm{r}_i+  \cos \bm{Q}_2 \cdot \bm{r}_i \\
    \cos \bm{Q}_1 \cdot \bm{r}_i-  \cos \bm{Q}_2 \cdot \bm{r}_i \\
    a_z (\sin \bm{Q}_1\cdot \bm{r}_i+\sin \bm{Q}_2\cdot \bm{r}_i)+ \tilde{M}_z
          \end{array}
  \right)^{\rm T}.
\end{align}
This state is also regarded as the vortex-antivortex crystal, where the antivortices are found around the skyrmion core with $S_i^z \simeq -1$ and vortices are found in the region between the neighboring skyrmion cores in the left panel in Fig.~\ref{fig:spin}(b). 
The core positions of the vortices and antivortices are locked at the center of the square plaquette reflecting the discrete lattice structure~\cite{Hayami_PhysRevResearch.3.043158}. 
Since the $z$ spin moment around the vortices (antivortices) points along the $+z$ ($-z$) direction, both the vortices and antivortices give the positive scalar chirality.
It is noted that there are other antivortices with $S_i^z>0$ in the region surrounded by the four antivortices with $S_i^z<0$ and four vortices with $S_i^z>0$, which contributes to the negative scalar chirality. 
As the contributions of the scalar chirality from the vortices and antivortices are different from each other, the scalar chirality in the whole system is not cancelled out in contrast to the double-$Q$ CS I state. 
The real-space scalar chirality configuration is shown in the middle left panel of Fig.~\ref{fig:spin}(b). 
By calculating the skyrmion number $n_{\rm sk}$ from the spin and scalar chirality configurations, one obtains a quantized skyrmion number of one, i.e., $n_{\rm sk}=1$, which indicates that the obtained state corresponds to the anti SkX. 
It is noted that the SkX with $n_{\rm sk}=-1$, which is expressed by reversing the sign of $\cos \bm{Q}_2 \cdot \bm{r}_i$ in Eq.~(\ref{eq:2QSkX}), is also obtained in the simulations with different initial spin configurations, since the energies between the SkX and anti SkX are degenerate in the present model.

\begin{figure}[t!]
\begin{center}
\includegraphics[width=1.0 \hsize ]{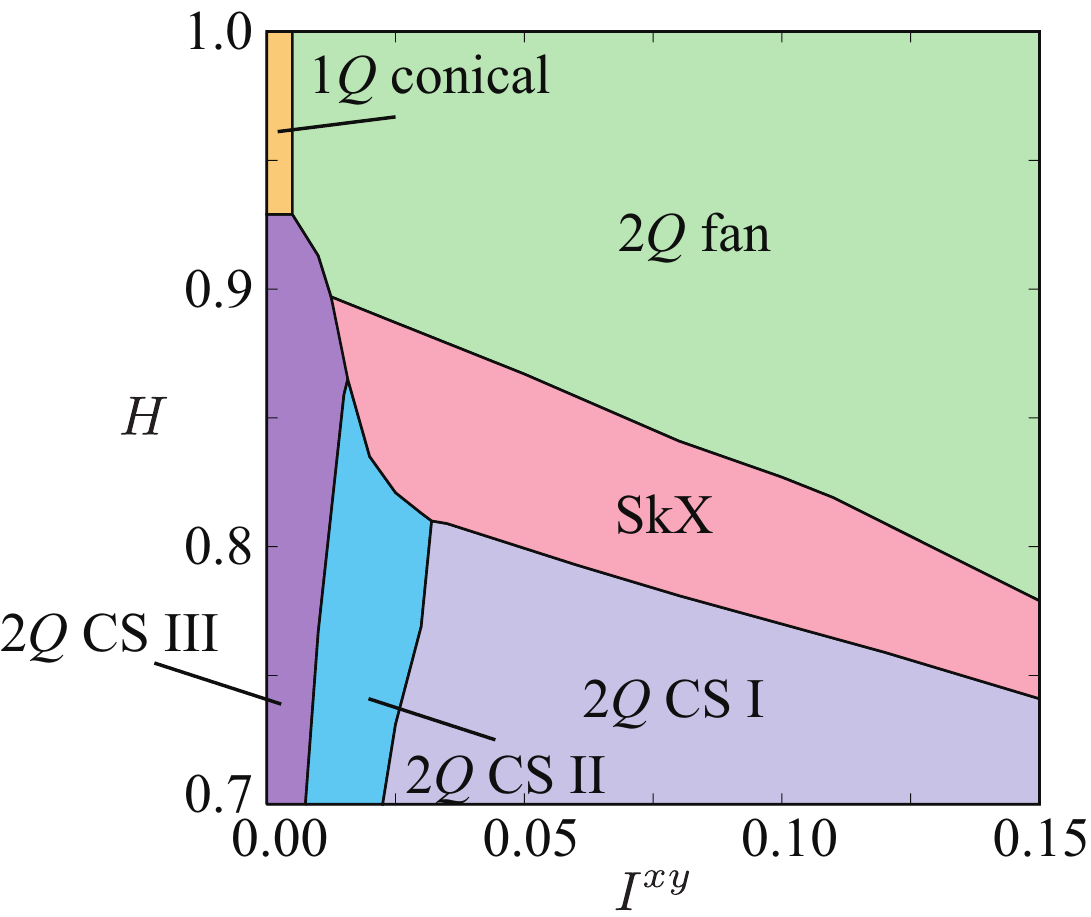} 
\caption{
\label{fig:Ani_PD}
Phase diagram in the plane of $I^{xy}$ and $H$ at $K=0.2$, $I^{z}=1.2$, and $I^v=0$ obtained by the simulated annealing. 
1$Q$ (2$Q$) represents the single-$Q$ (double-$Q$) state. 
CS represents the chiral stripe state. 
}
\end{center}
\end{figure}

\begin{figure*}[tb!]
\begin{center}
\includegraphics[width=1.0 \hsize ]{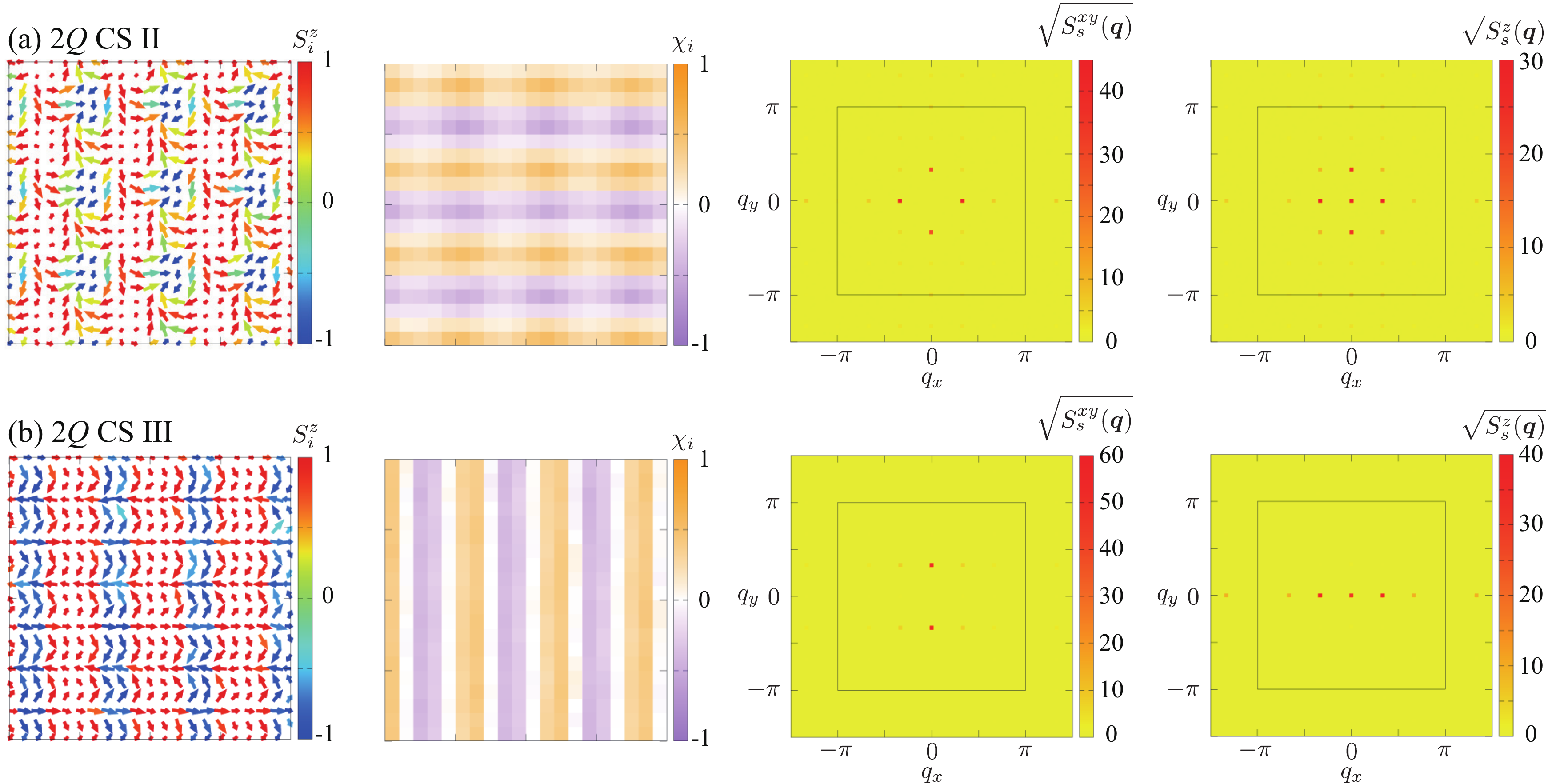} 
\caption{
\label{fig:spin2}
Left: Snapshots of the spin configurations in (a) the 2$Q$ chiral stripe (CS) II state for $I^{xy}=0.02$ and (b) the 2$Q$ CS III state for $I^{xy}=0.01$ at $H=0.8$, $I^z=1.2$, and $I^v=0$. 
The direction and the color of the arrows represent the $xy$ and $z$ components of the spin moment, respectively. 
Middle left: the scalar chirality $\chi_i$ calculated from the left panel. 
Middle right and right: The square root of the $xy$ and $z$ components of the spin structure factor. 
Black squares represent the first Brillouin zone.
}
\end{center}
\end{figure*}

While further increasing $H$, the SkX turns into the double-$Q$ fan state in Fig.~\ref{fig:PD}. 
The real-space spin and scalar chirality configurations are shown in the left and middle left panels of Fig.~\ref{fig:spin}(c), respectively. 
The $xy$ spin component is similar to that in the SkX in Fig.~\ref{fig:spin}(b), while there is a difference of the $z$ spin component; 
in the double-$Q$ fan state, there are almost no modulations in the $z$ spin component shown in the right panel of Fig.~\ref{fig:spin}(c). 
Indeed, there are double-$Q$ peaks at $\bm{Q}_1$ and $\bm{Q}_2$ in the $xy$ component of the spin structure factor, while there is almost no peak for $\bm{q}\neq 0$ in the $z$ component of the spin structure factor, as shown in the right two panels of Fig.~\ref{fig:spin}(c) (There are small intensities at $2\bm{Q}_1$ and $2\bm{Q}_2$ in the $z$ component).  
Reflecting such a spin configuration, the contributions of the scalar chirality from the vortex and antivortex become equivalent, which results in the cancellation of the scalar chirality in the whole system [the middle left panel of Fig.~\ref{fig:spin}(c)]. 
The expression of the spin configuration in the double-$Q$ fan state is represented by 
\begin{align}
\label{eq:2Qfan}
\bm{S}_i \propto 
\left(
    \begin{array}{c}
    \cos \bm{Q}_1 \cdot \bm{r}_i+  \cos \bm{Q}_2 \cdot \bm{r}_i \\
    \cos \bm{Q}_1 \cdot \bm{r}_i-  \cos \bm{Q}_2 \cdot \bm{r}_i \\
     \tilde{M}_z
          \end{array}
  \right)^{\rm T}. 
\end{align}
This spin configuration corresponds to that in the SkX in Eq.~(\ref{eq:2QSkX}) when taking $a_z=0$.

We show the $H$ dependence of the magnetization $M_z$ and the scalar chirality $(\chi_0)^2$ at $K=0.2$ in Fig.~\ref{fig:Mq}(a). 
There are clear jumps of $M_z$ when the phase transitions occur at $H \simeq 0.78$ and $H \simeq 0.86$. 
This indicates the first-order phase transition between the SkX and the other two double-$Q$ states. 
One also finds that only the SkX shows a nonzero scalar chirality $(\chi_0)^2$. 
Figure~\ref{fig:Mq}(b) shows the $H$ dependence of $m^\mu_{\bm{Q}_1}$ and $m^\mu_{\bm{Q}_2}$ for $\mu=x,y,z$. 
The double-$Q$ CS I state has anisotropic double-$Q$ components ($m^\mu_{\bm{Q}_1} \neq m^\mu_{\bm{Q}_2}$), while the SkX and double-$Q$ fan states has isotropic double-$Q$ components ($m^\mu_{\bm{Q}_1} = m^\mu_{\bm{Q}_2}$). 
The $x$- and $y$-spin components in all the states are equivalent with each other due to the nature of $I^{xy}$, i.e., $m^x_{\bm{Q}_1} = m^y_{\bm{Q}_1}$ and $m^x_{\bm{Q}_2} = m^y_{\bm{Q}_2}$.

Next, we discuss the stability of the SkX while changing $I^{xy}$. 
For that purpose, we fix $K=0.2$ and construct the low-temperature phase diagram in the $I^{xy}$-$H$ plane with $\theta=0$. 
Figure~\ref{fig:Ani_PD} shows the phase diagram obtained by the simulated annealing. 
We here focus on the magnetic field region where the SkX is stabilized. 
As shown in Fig.~\ref{fig:Ani_PD}, the SkX appears in the region for $I^{xy} \gtrsim 0.013$, which means that small but nonzero $I^{xy}$ is important to stabilize the SkX. 
The stable field range of the SkX becomes the largest around $I^{xy} \simeq 0.03$.
The high-field phase of the SkX is always the double-$Q$ fan state. 
Meanwhile, the low-field phases of the SkX are different depending on $I^{xy}$: the double-$Q$ CS I state for $I^{xy} \gtrsim 0.037$, the double-$Q$ CS II state for $0.016 \lesssim I^{xy} \lesssim 0.037$, and the double-$Q$ CS III state for $0.013 \lesssim I^{xy} \lesssim 0.016$. 

We show the spin and scalar chirality configurations in the double-$Q$ CS II and III states in Fig.~\ref{fig:spin2}. 
The double-$Q$ CS II state is characterized by the anisotropic double-$Q$ structure, which is similar to the double-$Q$ CS I state in Fig.~\ref{fig:spin}(a). 
Although the $xy$-spin component is similar between the two states as shown in the middle right panel of Figs.~\ref{fig:spin}(a) and \ref{fig:spin2}(a), their difference is found in the $z$-spin component: 
The double-$Q$ peaks appear at $\bm{Q}_1$ and $\bm{Q}_2$ in the double-$Q$ CS II state, while only the single-$Q$ peak appears in the double-$Q$ CS I state, as shown in the right panel of Figs.~\ref{fig:spin}(a) and \ref{fig:spin2}(a). 
Reflecting the double-$Q$ structure in both $xy$ and $z$ spin components, the real-space spin configuration in the left panel of Fig.~\ref{fig:spin2}(a) seems to be similar to that in the SkX in Fig.~\ref{fig:spin}(b). 
However, there is no net scalar chirality in this state.  
The scalar chirality density wave occurs for $\bm{q} \neq 0$, as shown in the middle left panel of Fig.~\ref{fig:spin2}(a); the dominant peak is found in $\bm{Q}_2$ and the subdominant peaks are found in $\bm{Q}_1+\bm{Q}_2$ and $2\bm{Q}_2$.

In the small $I^{xy}$ region, the double-$Q$ CS III state appears in the phase diagram in Fig.~\ref{fig:Ani_PD}. 
The spin configuration in this state is characterized by the superposition of the horizontal spiral wave along the $\bm{Q}_2$ direction and the sinusoidal wave along the $\bm{Q}_1$ direction, as shown in the right two panels of Fig.~\ref{fig:spin2}(b). 
Such a superposition is also found in the real-space spin configuration, as shown in the left panel of Fig.~\ref{fig:spin2}(b). 
This state also accompanies the chirality density wave along the $\bm{Q}_1$ direction, as shown in the middle left panel of Fig.~\ref{fig:spin2}(b). 
When the magnetic field is increased in the small $I^{xy}$ region, the sinusoidal $\bm{Q}_1$ component vanishes, and then, the state turns into the single-$Q$ conical state.

From the above phase diagrams in Figs.~\ref{fig:PD} and \ref{fig:Ani_PD}, one finds that the interplay between the biquadratic interaction $K$ and the anisotropic interaction $I^{xy}$ originating from the vertical mirror symmetry breaking plays an important role in stabilizing the square SkX. 
The result indicates that the set of large $K$ and moderate $I^{xy}$ is preferable to obtain the robust square SkX. 
Moreover, the identification of the low-field phase in experiments provides information about the magnitude of $I^{xy}$; the emergence of the 2$Q$ CS I (III) state corresponds to the large (small) anisotropic interaction.

\subsection{Helicity locking of skyrmion crystal}
\label{sec:Helicity locking of skyrmion crystal}

\begin{figure}[t!]
\begin{center}
\includegraphics[width=1.0 \hsize ]{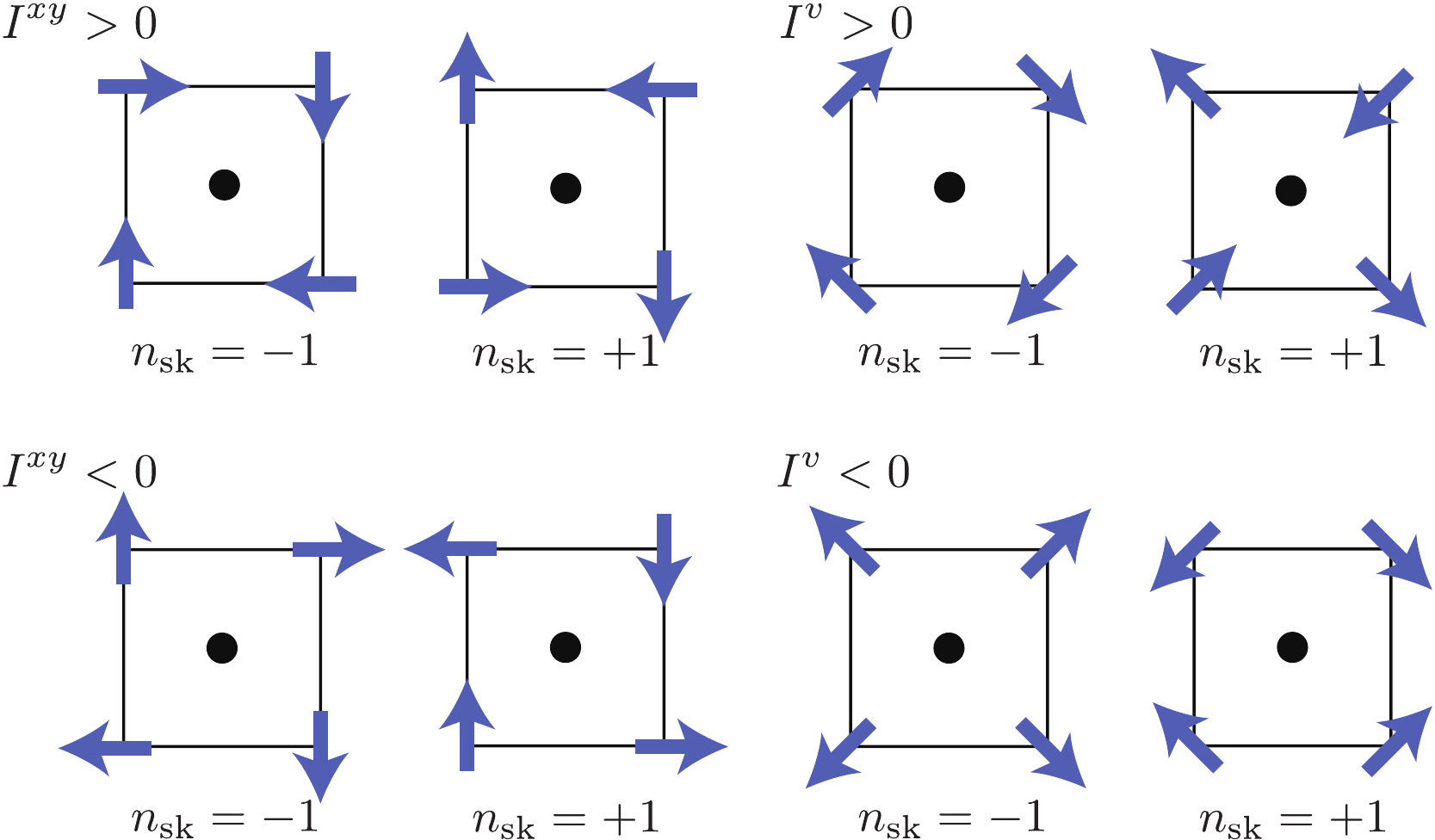} 
\caption{
\label{fig:Table_IxyIa}
Schematic inplane spin configurations (blue arrows) around the skyrmion core (black circles) stabilized in the presence  of (left panel) $I^{xy}$ and (right panel) $I^v$. 
The skyrmion number $n_{\rm sk}$ in each spin texture is also shown. 
}
\end{center}
\end{figure}

\begin{figure}[t!]
\begin{center}
\includegraphics[width=1.0 \hsize ]{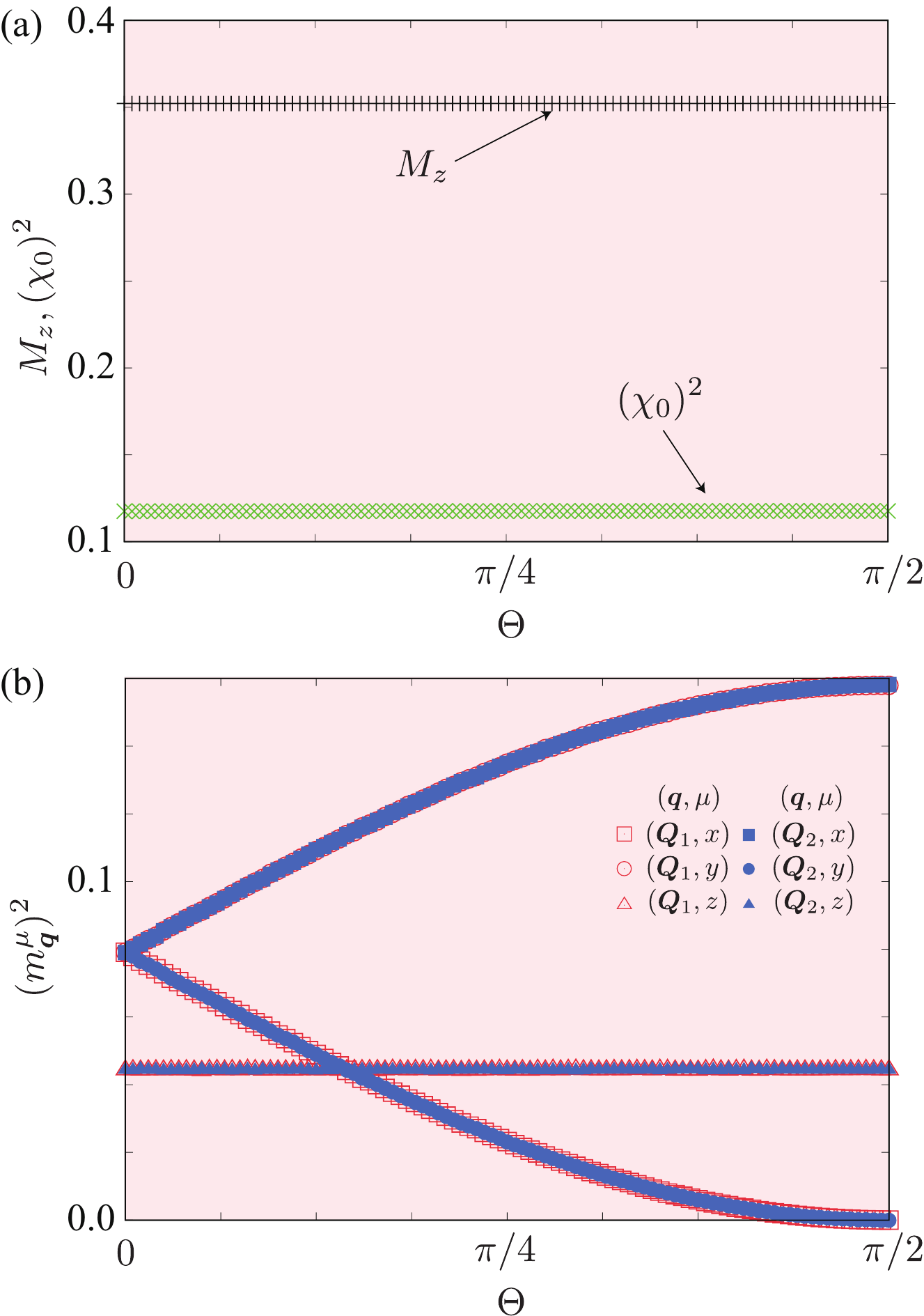} 
\caption{
\label{fig:Mq_helicity}
$\Theta$ dependences of (a) $M_z$, $(\chi_0)^2$, (b) $m^\mu_{\bm{Q}_1}$, and $m^\mu_{\bm{Q}_2}$ for $\mu=x,y,z$ at $K=0.3$, $I^z=1.2$, $I^{a}=0.05$, and $H=0.75$.  
}
\end{center}
\end{figure}

\begin{figure}[t!]
\begin{center}
\includegraphics[width=1.0 \hsize ]{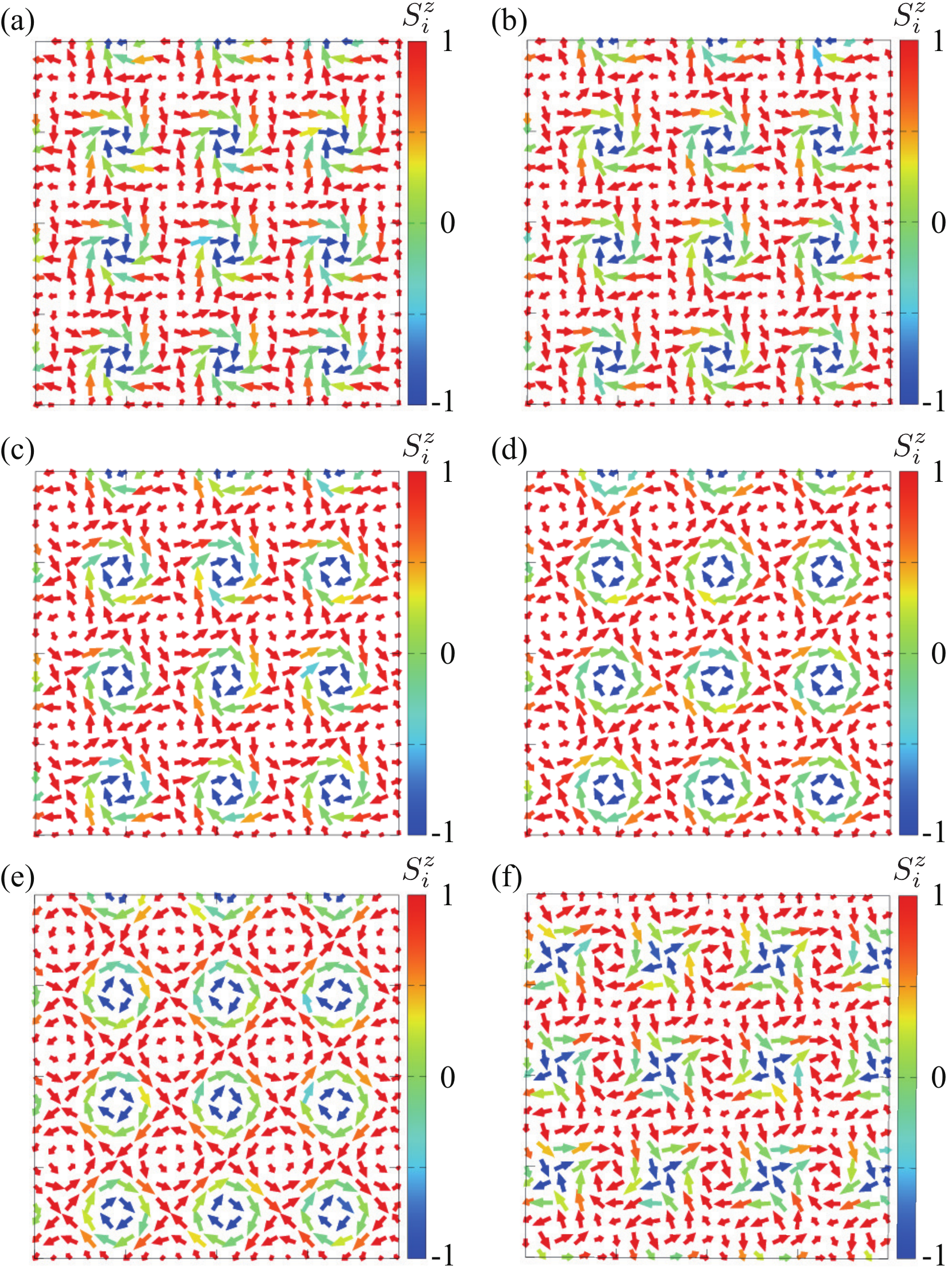} 
\caption{
\label{fig:spin_IxyIa}
(a-e) Snapshots of the spin configurations in the SkX with $n_{\rm sk}=-1$ for (a) $\Theta=\pi/200$, (b) $\Theta=\pi/8$, (c) $\Theta=\pi/4$, (d) $\Theta=3\pi/8$, and (e) $\Theta=\pi/2$ at $I^a=0.05$, $K=0.3$, $I^z=1.2$, and $H=0.75$.
(f) Snapshots of the spin configurations in the SkX with $n_{\rm sk}=+1$ obtained by the simulations when starting from a different random spin configuration in (c). 
}
\end{center}
\end{figure}

In this section, we discuss the helicity of the SkX. 
As discussed in Sec.~\ref{sec:Skyrmion crystal in an out-of-plane field}, two types of the SkXs are obtained in the simulations, which are characterized by the different skyrmion number $n_{\rm sk}=\pm 1$. 
This means that there is a degeneracy in terms of the vorticity of the skyrmion core. 
On the other hand, the helicity of the skyrmion core is fixed at $\pi/4$ or $-3\pi/4$ for the SkX with $n_{\rm sk}=-1$ and at $-\pi/4$ and $3\pi/4$ for the anti SkX with $n_{\rm sk}=1$, where the helicity is defined by the angle between $\bm{S}_i$ and $\bm{R}_i$ [$\bm{R}_i=(X_i, Y_i)$ 
is the position vector measured from the skyrmion core (center of the square plaquette)]. 
For example, the snapshot in the left panel of Fig.~\ref{fig:spin}(b) corresponds to the helicity $-\pi/4$ and $3\pi/4$. 
When the sign of $I^{xy}$ is reversed, the helicity takes opposite values; the SkX with $n_{\rm sk}=-1$ has $-\pi/4$ or $3\pi/4$ and the anti SkX with $n_{\rm sk}=1$ has the helicity $\pi/4$ and $-3\pi/4$. 
The spin textures around the skyrmion core in the presence of $I^{xy}$ are schematically summarized in the left panel of Fig.~\ref{fig:Table_IxyIa}. 

Meanwhile, another anisotropic interaction $I_v$ that arises from the discrete fourfold rotational symmetry around the $z$ axis also fixes the helicity of the skyrmion core in a different manner~\cite{Hayami_PhysRevB.103.024439}. 
In the absence of $I^{xy}$ where the lattice symmetry reduces to the $D_{\rm 4h}$ symmetry, the helicity of the skyrmion core is fixed at $\pi/2$ or $-\pi/2$ for the SkX with $n_{\rm sk}=-1$ and at $0$ and $\pi$ for the anti SkX with $n_{\rm sk}=1$ for $I^v>0$. 
In the case of $I^v <0$, the tendency is opposite; the helicity of the skyrmion core is fixed at $0$ or $\pi$ for the SkX with $n_{\rm sk}=-1$ and at $\pi/2$ and $-\pi/2$ for the anti SkX with $n_{\rm sk}=1$. 
The spin configurations around the skyrmion core for $I^v \neq 0$ and $I^{xy}=0$ are schematically shown in the right panel of Fig.~\ref{fig:Table_IxyIa}.

The different tendency with respect to the helicity locking by $I^{xy}$ and $I^v$ indicates that the helicity in real materials is determined by taking into account both $I^{xy}$ and $I^v$. 
Such a situation naturally happens in the $C_{\rm 4 h}$ point group system, as discussed in Sec.~\ref{sec:Effective spin model}. 
In the following, we show that the helicity of the SkX is determined by the ratio of two anisotropic interactions, $I^{xy}$ and $I^v$. 
To demonstrate that, we denote two anisotropic interactions as $(I^{xy}, I^{v})=I^a (\cos \Theta, \sin \Theta)$ and change $\Theta$ for fixed $I^a$. 

The results of $M_z$, $(\chi_0)^2$, and $(m^\mu_{\bm{Q}_\eta})^2$ at $K=0.3$, $I^z=1.2$, $I^a=0.05$, and $H=0.75$ against $\Theta$ are shown in Fig.~\ref{fig:Mq_helicity}.
The data in Figs.~\ref{fig:Mq_helicity}(a) and \ref{fig:Mq_helicity}(b) clearly represent no $\Theta$ dependence of the quantities irrelevant of the helicity, $M_z$, $(\chi_0)^2$, and $(m^z_{\bm{Q}_\eta})^2$. 
Meanwhile, $(m^x_{\bm{Q}_\eta})^2$ and $(m^y_{\bm{Q}_\eta})^2$ in Fig.~\ref{fig:Mq_helicity}(b) shows a $\Theta$ dependence to smoothly connect the result at $I^v=0$ in Sec.~\ref{sec:Skyrmion crystal in an out-of-plane field} and that at $I^{xy}=0$ in Ref.~\onlinecite{Hayami_PhysRevB.103.024439}. 
We show the snapshots obtained by the simulated annealing for several $\Theta$ in Figs.~\ref{fig:spin_IxyIa}(a)-\ref{fig:spin_IxyIa}(e).  
For $\Theta \simeq 0$, the SkX with $n_{\rm sk}=-1$ and helicity $-3\pi/4$ is realized, as shown in Fig.~\ref{fig:spin_IxyIa}(a). 
While increasing $\Theta$, i.e., $I^v$, the helicity gradually changes from $-3\pi/4$ to $-\pi/2$, as shown in Figs.~\ref{fig:spin_IxyIa}(b)-\ref{fig:spin_IxyIa}(e). 
This result indicates a correspondence between the helicity of the skyrmion and the ratio of $I^{xy}$ and $I^v$. 
Thus, one can estimate the ratio of $I^{xy}$ and $I^v$, once the helicity of the skyrmion is identified and vice versa. 
Although it is difficult to estimate the helicity from the real-space observation owing to its resolution, the measurement of intensities at magnetic moments with $\bm{Q}_1$ and $\bm{Q}_2$ components as shown in Fig.~\ref{fig:Mq_helicity}(b) by using the resonant x-ray scattering with the polarization analysis can be performed.
Indeed, the change of the skyrmion helicity at the surface has been observed in the chiral magnet Cu$_2$OSeO$_3$~\cite{PhysRevB.87.094424,PhysRevLett.120.227202}.

It is noted that the degeneracy between the SkXs with $n_{\rm sk}=\pm 1$ still remains even in the presence of both $I^{xy}$ and $I^v$. 
We show the snapshot of the SkX with $n_{\rm sk}=+1$ for $\Theta=\pi/4$, which is obtained by the simulations starting from a different random spin configuration, in Fig.~\ref{fig:spin_IxyIa}(f). 
In this case, the real-space spin configuration around the skyrmion core is described by the superposition of the spin configuration shown at $I^{xy}>0$ and $n_{\rm sk}=+1$ and that shown at $I^v>0$ and $n_{\rm sk}=+1$ in Fig.~\ref{fig:Table_IxyIa}.

\subsection{Relation to odd-parity multipoles}
\label{sec:Relation to odd-parity multipoles}

The emergence of the skyrmion spin texture in centrosymmetric magnets breaks both time-reversal and spatial inversion symmetries spontaneously.  
Meanwhile, as the SkXs with the different helicity are categorized into the different irreducible representation under the point group, different physical responses can be expected. 
In this section, we present the helicity-dependent physical phenomena from the viewpoint of odd-parity multipoles with the spatial-inversion odd, since the concept of multipoles gives a systematic understanding of transport properties and multiferroic phenomena~\cite{Hayami_PhysRevB.98.165110,suzuki2018first,Watanabe_PhysRevB.98.245129,Yatsushiro_PhysRevB.104.054412}. 
The relevant odd-parity magnetic and magnetic toroidal multipoles that can be a source of the linear magnetoelectric effect are discussed in Sec.~\ref{sec:Odd-parity magnetic and magnetic toroidal multipoles} and the relevant odd-parity electric and electric toroidal multipoles that can be a source of the antisymmetric spin splitting in the band structure and the Edelstein effect are discussed in Sec.~\ref{sec:Odd-parity electric and electric toroidal multipoles}. 

\begin{figure}[t!]
\begin{center}
\includegraphics[width=1.0 \hsize ]{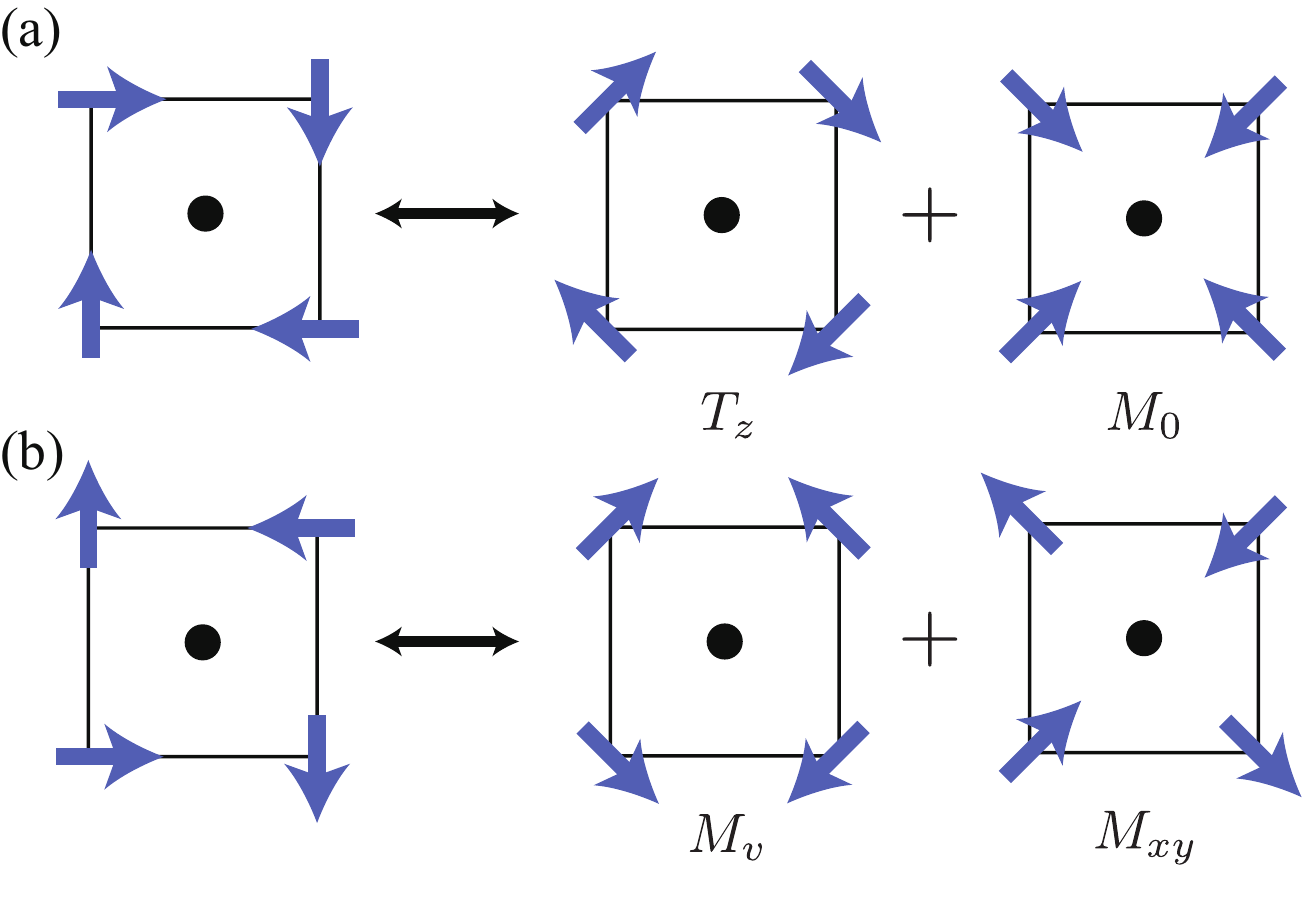} 
\caption{
\label{fig:Multipole}
The correspondence between the skyrmion spin textures and the odd-parity multipoles. 
$M_0$ represents the magnetic monopole, $T_z$ represents the magnetic toroidal dipole, and $M_v$ and $M_{xy}$ represent the magnetic quadrupoles.  
}
\end{center}
\end{figure}

\subsubsection{Odd-parity magnetic and magnetic toroidal multipoles}
\label{sec:Odd-parity magnetic and magnetic toroidal multipoles}

The SkXs with the different helicity have the different spin configurations around the skyrmion core, as shown in Fig.~\ref{fig:Table_IxyIa}. 
We here classify the spin configurations with the different helicity on the basis of magnetic and magnetic toroidal multipoles. 
As the magnetic multipoles are characterized by an axial tensor and the magnetic toroidal multipoles are characterized by a polar tensor, the even-rank magnetic multipoles and the odd-rank magnetic toroidal multipoles correspond to the odd-parity multipoles in terms of spatial inversion symmetry~\cite{dubovik1975multipole,Hayami_PhysRevB.98.165110,Yatsushiro_PhysRevB.104.054412}. 
In other words, these odd-parity magnetic and magnetic toroidal multipoles are fundamental multipole degrees of freedom in the absence of both spatial inversion and time-reversal symmetries, which have often been discussed in the field of multiferroics~\cite{kopaev2009toroidal,Spaldin_0953-8984-20-43-434203,Hayami_PhysRevB.90.024432,Watanabe_PhysRevB.96.064432,thole2018magnetoelectric,Gao_PhysRevB.98.060402,Shitade_PhysRevB.98.020407, thole2020concepts,Bhowal_PhysRevResearch.3.033185,Hayami_PhysRevB.104.045117} including the SkXs~\cite{seki2012observation,white2012electric,okamura2013microwave,Mochizuki_PhysRevB.87.134403,tokura2014multiferroics,mochizuki2015dynamical,Christensen_PhysRevX.8.041022,Gobel_PhysRevB.99.060406}.

\begin{table}[t!]
\centering
\caption{
Active odd-parity magnetic and magnetic toroidal multipoles (OMP) and nonzero magnetoelectric tensor components $\alpha_{\mu\nu}$ for a set of $(n_{\rm sk}, I_v, I_{xy})$.
The sign in the columns $I_v$ and $I_{xy}$ represents the sign of their anisotropic form factors.  
\label{table:mp}}
\vspace{2mm}
\begin{tabular}{cccccccccccccccccccc}
\hline\hline
$n_{\rm sk}$ & $I_v$ &  $I_{xy}$ & OMP & $\alpha_{\mu\nu}$\\ \hline 
$-1$ & + & 0 & $T_z$ & $\alpha_{xy}=-\alpha_{yx}$ \\
$+1$ & + & 0 & $M_{xy}$& $\alpha_{xy}=\alpha_{yx}$\\ 
$-1$ & - & 0 & $M_0$& $\alpha_{xx}=\alpha_{yy}$ \\
$+1$ & - & 0 & $M_v$&$\alpha_{xx}=-\alpha_{yy}$\\ \hline
$-1$ & 0 & + &  $M_0+T_z$  & $\alpha_{xx}=\alpha_{yy}=\alpha_{xy}=-\alpha_{yx}$  \\
$+1$ & 0 & + & $M_v+M_{xy}$ & $\alpha_{xx}=-\alpha_{yy}=\alpha_{xy}=\alpha_{yx}$\\ 
 $-1$ & 0 & - & $M_0-T_z$ & $\alpha_{xx}=\alpha_{yy}=-\alpha_{xy}=\alpha_{yx}$\\
$+1$ & 0 & - & $M_v-M_{xy}$ & $\alpha_{xx}=-\alpha_{yy}=-\alpha_{xy}=-\alpha_{yx}$\\ \hline
$-1$ & $\pm$ & $\pm$ &  $M_0, T_z$  & $\alpha_{xx}=\alpha_{yy}, \alpha_{xy}=-\alpha_{yx}$  \\
$+1$ & $\pm$ & $\pm$ & $M_v, M_{xy}$ & $\alpha_{xx}=-\alpha_{yy}, \alpha_{xy}=\alpha_{yx}$\\ 
\hline\hline
\end{tabular}
\end{table}

By using cluster multipole theory for the four-site cluster around the skyrmion core~\cite{Suzuki_PhysRevB.95.094406,Suzuki_PhysRevB.99.174407}, one finds that four types of odd-parity multipoles can become active, whose expressions are given by 
\begin{align}
\label{eq:M0}
M_0&= \sum_i \bm{R}_i \cdot \bm{S}_i, \\
\label{eq:Tz}
T_z&= \sum_i (\bm{R}_i \times \bm{S}_i)^z,  \\
\label{eq:Mv}
M_v&= \sum_i X_i S^x_i - Y_i S^y_i, \\
\label{eq:Mxy}
M_{xy}&= \sum_i X_i S^y_i + Y_i S^x_i, 
\end{align}
where $M_0$ represents the rank-0 magnetic monopole, $T_z$ represent the rank-1 magnetic toroidal dipole, and $M_v$ and $M_{xy}$ represent the rank-2 magnetic quadrupoles, where we omit the irrelevant numerical coefficient~\cite{comment_Mu}. 
$M_0$ and $T_z$ are induced when the skyrmion core with $n_{\rm sk}=-1$, while $M_v$ and $M_{xy}$ are induced when the skyrmion core with $n_{\rm sk}=+1$. 
The schematic spin configurations for $M_0$, $T_z$, $M_v$, and $M_{xy}$ are shown in Fig.~\ref{fig:Multipole}. 
Under the point group $D_{\rm 4h}$ ($C_{\rm 4h}$), the irreducible representations of $M_0$, $T_z$, $M_v$, and $M_{xy}$ correspond to ${\rm A}_{1u}$, ${\rm A}_{2u}$, ${\rm B}_{1u}$, and ${\rm B}_{2u}$ (${\rm A}_{u}$, ${\rm A}_{u}$, ${\rm B}_{u}$, and ${\rm B}_{u}$), respectively. 

In the case of $I^v \neq 0$ and $I^{xy}=0$, one of four odd-parity multipoles is activated depending on $n_{\rm sk}$ and helicity, as shown in Fig.~\ref{fig:Table_IxyIa}. 
Meanwhile, for $I^v = 0$ and $I^{xy}\neq 0$, the spin texture is represented by the linear combination of $T_z$ and $M_0$ ($M_v$ and $M_{xy}$) for the SkX with $n_{\rm sk}=-1$ ($n_{\rm sk}=+1$), as shown in Fig.~\ref{fig:Multipole}. 
As the number of active multipoles is related to the nonzero response tensor as described below, it is expected that the SkXs under the point group $C_{\rm 4h}$ exhibit rich physical phenomena than those under $D_{\rm 4h}$. 
This argument is consistent with the symmetry analysis based on the point group; the irreducible representation ${\rm A}_{1g/u}$ and ${\rm A}_{2g/u}$ (${\rm B}_{1g/u}$ and ${\rm B}_{2g/u}$) under $D_{\rm 4h}$ belong to the same irreducible representation ${\rm A}_{g/u}$ (${\rm B}_{g/u}$) under $C_{\rm 4h}$, which means that $M_0$ and $T_z$ ($M_v$ and $M_{xy}$) are not distinguished from the symmetry viewpoint. 

The active odd-parity magnetic and magnetic toroidal multipoles are closely related to the linear magnetoelectric effect, where the magnetization $M_{\mu}$ is induced by the electric field $E_{\nu}$ represented by $M_{\mu}=\sum_{\nu}\alpha_{\mu\nu}E_{\nu}$ for $\mu,\nu=x,y$ (Here and hereafter, we only consider the $\mu,\nu=x,y$ components for simplicity).  
The nonzero magnetoelectric tensor $\alpha_{\mu\nu}$ has a correspondence with four odd-parity multipoles in Eqs.~(\ref{eq:M0})-(\ref{eq:Mxy}) as
\begin{align}
\label{eq:alphaxx}
\alpha_{xx}&=M_0+M_v, \\
\label{eq:alphayy}
\alpha_{yy}&=M_0-M_v, \\
\label{eq:alphaxy}
\alpha_{xy}&=M_{xy}+T_z,\\
\label{eq:alphayx}
\alpha_{yx}&=M_{xy}-T_z. 
\end{align}
Active $M_0$ and $M_v$ give rise to the longitudinal magnetoelectric effect, while active $T_z$ and $M_{xy}$ lead to the transverse one. 
From the correspondence between the active odd-parity multipoles and the anisotropic form factors as discussed above, one finds that $\alpha_{\mu\nu}$ has one independent component for $I_{v}\neq 0$ and $I_{xy}=0$ or $I_{xy} \neq 0$ and $I_{v}=0$ 
, whereas $\alpha_{\mu\nu}$ has two independent components for $I_{v}\neq 0$ and $I_{xy}\neq 0$. 
The conditions in each set of $(n_{\rm sk}, I_v, I_{xy})$ to induce nonzero $(M_0, T_z, M_v, M_{xy})$ and $(\alpha_{xx}, \alpha_{yy}, \alpha_{xy}, \alpha_{yx})$ are summarized in Table~\ref{table:mp}.

\subsubsection{Odd-parity electric and electric toroidal multipoles}
\label{sec:Odd-parity electric and electric toroidal multipoles}

\begin{table}[t!]
\centering
\caption{
Correspondence between active odd-parity multipoles among magnetic, magnetic toroidal, electric, and electric toroidal multipoles under the square SkX. 
The functional form of the antisymmetric spin splitting in momentum space $k_\mu\sigma_\nu$ and the magneto-current tensor components $\tilde{\alpha}_{\mu\nu}$ are also shown. 
\label{table:mp2}}
\vspace{2mm}
\begin{tabular}{cccccccccccccccccccc}
\hline\hline
correspondence & $k_{\mu}\sigma_{\nu}$ & $\tilde{\alpha}_{\mu\nu}$ \\ \hline 
$M_0 \leftrightarrow Q_z$ & $k_x \sigma_y - k_y \sigma_x$ & $\tilde{\alpha}_{xy}=-\tilde{\alpha}_{yx}$ \\
$T_z  \leftrightarrow -G_0$ & $-k_x \sigma_x - k_y \sigma_y$ & $\tilde{\alpha}_{xx}=\tilde{\alpha}_{yy}$ \\
$M_{v} \leftrightarrow -G_{xy}$ & $-k_x \sigma_y - k_y \sigma_x$ & $\tilde{\alpha}_{xy}=\tilde{\alpha}_{yx}$ \\
$M_{xy} \leftrightarrow G_v$ & $k_x \sigma_x - k_y \sigma_y$ & $\tilde{\alpha}_{xx}=-\tilde{\alpha}_{yy}$ \\
\hline\hline
\end{tabular}
\end{table}

Considering that there is a uniform magnetization in the SkX phase, odd-parity electric and electric toroidal multipoles become active in addition to odd-parity magnetic and magnetic toroidal multipoles owing to the breaking of the product symmetry of spatial inversion and time-reversal symmetries. 
Here, the electric (electric toroidal) multipoles are characterized by a polar (axial) tensor with the time-reversal even~\cite{dubovik1975multipole,Hayami_PhysRevB.98.165110,Yatsushiro_PhysRevB.104.054412}. 
Specifically, the odd-rank electric multipoles and the even-rank electric toroidal multipoles correspond to the odd-parity multipoles. 
For example, the electric dipole is active in the polar systems like the Rashba system, and the electric toroidal monopole is active in the chiral systems like the Weyl system. 
Recently, electric toroidal quadrupole ordering has been suggested in Cd$_2$Re$_2$O$_7$~\cite{hiroi2017pyrochlore,Matteo_PhysRevB.96.115156,Hayami_PhysRevLett.122.147602} and CeCoSi~\cite{yatsushiro2020odd}. 
The active odd-parity electric and electric toroidal multipoles can be a source of the antisymmetric spin splitting in the band structure and the Edelstein effect, as discussed below. 

From the symmetry viewpoint~\cite{Hayami_PhysRevB.98.165110,Yatsushiro_PhysRevB.104.054412}, we focus on four types of the electric and electric toroidal multipoles, which become active in the presence of $(M_0, T_z, M_{v}, M_{xy})$ under the magnetic field: the rank-1 electric dipole $Q_z$, the rank-0 electric toroidal monopole $G_0$, and the rank-2 electric toroidal quadrupoles $G_v, G_{xy}$~\cite{comment_Gu}. 
The correspondence between them is given by $M_0 \leftrightarrow Q_z$, $T_z \leftrightarrow -G_0$, $M_v \leftrightarrow -G_{xy}$, and $M_{xy} \leftrightarrow G_v$~\cite{Yatsushiro_PhysRevB.104.054412}. 
These four types of multipoles are related to the antisymmetric spin-split band structure as~\cite{Hayami_PhysRevB.98.165110} 
\begin{align}
\label{eq:G0}
G_0&= \bm{k} \cdot \bm{\sigma}, \\
\label{eq:Qz}
Q_z&=  (\bm{k} \times \bm{\sigma})^z,  \\
\label{eq:Qv}
G_v&= k_x \sigma_x - k_y \sigma_y, \\
\label{eq:Qxy}
G_{xy}&=  k_x \sigma_y + k_y \sigma_x, 
\end{align}
where $\bm{k}$ is the wave vector and $\bm{\sigma}$ is the spin. 
We consider the expression of $\bm{k} \to \bm{0}$ for simplicity.  
As the active multipoles depend on the helicity of the SkXs, the different types of the antisymmetric spin splitting occur according to the different helicity. 
Such a different $\bm{k}$-resolved spin polarization can be detected by the spin- and angle-resolved photoemission spectroscopy measurement. 
In other words, the spin- and angle-resolved photoemission spectroscopy measurement is one of the probe for the helicity through the $\bm{k}$-resolved spin polarization. 

In addition, the active odd-parity electric and electric toroidal multipoles lead to the Edelstein effect where the magnetization $M_\mu$ is induced by the electric current $J_\nu$ in metals: $M_{\mu}=\sum_{\nu}\tilde{\alpha}_{\mu\nu}J_{\nu}$ ($\tilde{\alpha}$ represents the magneto-current tensor). 
The nonzero $\tilde{\alpha}_{\mu\nu}$ is related with four odd-parity multipoles in Eqs.~(\ref{eq:G0})-(\ref{eq:Qxy}), which is obtained by replacing $(M, T)$ in Eqs.~(\ref{eq:alphaxx})-(\ref{eq:alphayx}) with $(G, Q)$~\cite{Hayami_PhysRevB.98.165110}: 
\begin{align}
\tilde{\alpha}_{xx}&=G_0+G_v, \\
\tilde{\alpha}_{yy}&=G_0-G_v, \\
\tilde{\alpha}_{xy}&=G_{xy}+Q_z,\\
\tilde{\alpha}_{yx}&=G_{xy}-Q_z. 
\end{align}
Similar to $(M_0, T_z, M_v, M_{xy})$, active $G_0$ and $G_v$ induce the longitudinal Edelstein effect, while active $Q_z$ and $G_{xy}$ induce the transverse one. 
The correspondence between four types of odd-parity multipoles and its relation to the antisymmetric spin splitting and the magneto-current tensor are summarized in Table~\ref{table:mp2}. 

It is noted that the present antisymmetric spin-split band structure in the SkX is caused by the magnetic phase transitions rather than the antisymmetric spin-orbit coupling in a noncentrosymmetric lattice structure. 
There, the noncollinear magnetic texture plays an important role in inducing the antisymmetric spin-split band structure~\cite{Hayami_PhysRevB.101.220403,Hayami_PhysRevB.102.144441,Yuan_PhysRevMaterials.5.014409}. 
Indeed, it was shown that the spin textures with the chiral-type bilinear spin product $\bm{S}_{\bm{q}} \times \bm{S}_{-\bm{q}}$, which becomes nonzero in the spiral spin texture, are related to the appearance of the antisymmetric spin splitting~\cite{hayami2022mechanism}.  
For example, the SkX spin texture in Eq.~(\ref{eq:2QSkX}), which possesses $M_{xy}+M_{v}$, has nonzero $\bm{S}_{\bm{Q}_1} \times \bm{S}_{-\bm{Q}_1}$ and $\bm{S}_{\bm{Q}_2} \times \bm{S}_{-\bm{Q}_2}$, which leads to $k_x \sigma_x - k_x \sigma_y$ and $-k_y \sigma_x - k_y \sigma_y$, respectively. 
In other words, the SkX shows the antisymmetric spin splitting in the form of $(k_x-k_y) \sigma_x + (-k_x-k_y) \sigma_y = G_v-G_{xy}$, which is consistent with the above multipole argument.

\subsection{Stability of skyrmion crystal in a rotated field}
\label{sec:Stability of skyrmion crystal in a rotated field}

\begin{figure}[t!]
\begin{center}
\includegraphics[width=1.0 \hsize ]{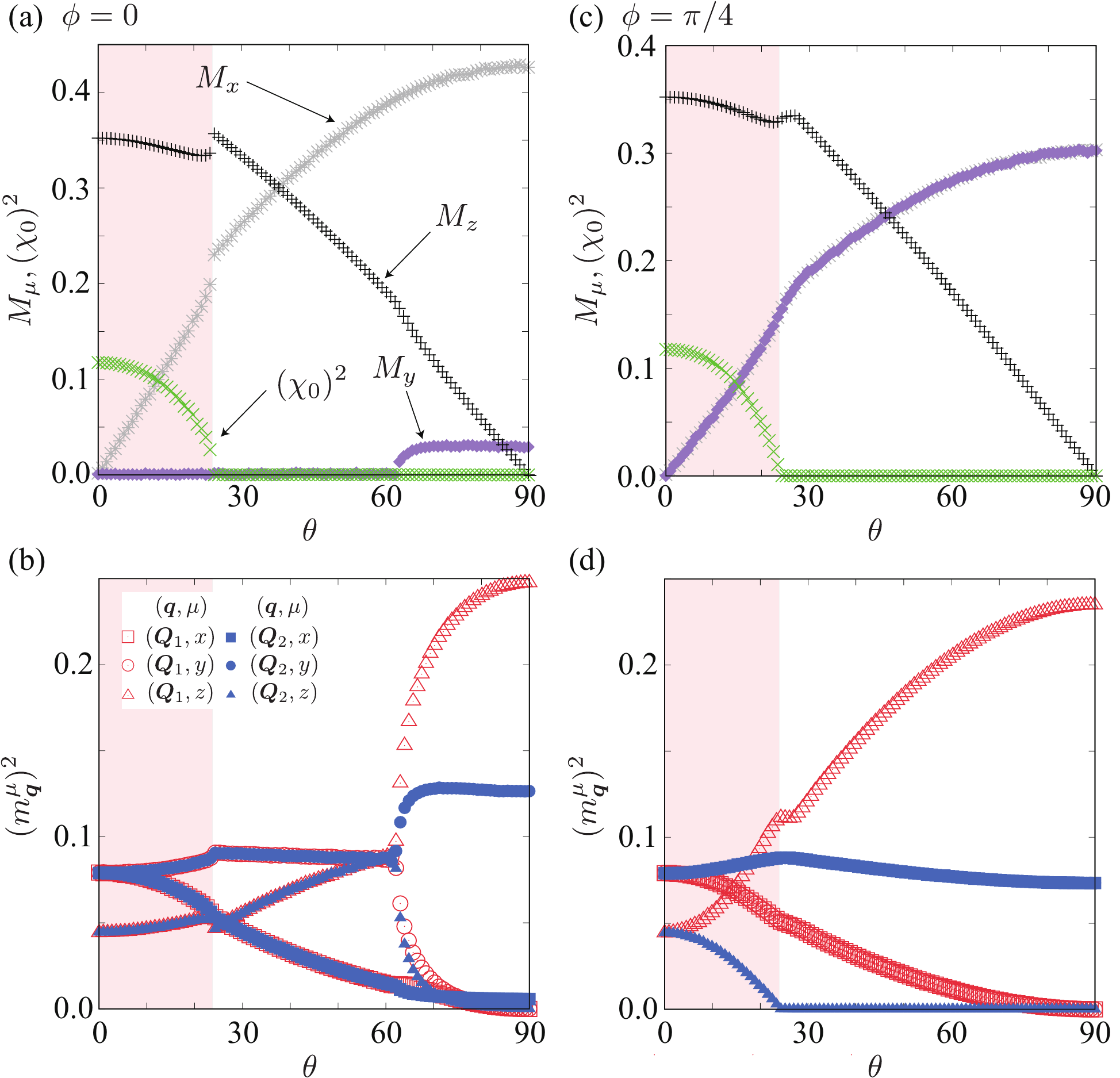} 
\caption{
\label{fig:Mq_rotH_1}
$\theta$ dependences of (a,c) $M_\mu$, $(\chi_0)^2$, (b,d) $m^\mu_{\bm{Q}_1}$, and $m^\mu_{\bm{Q}_2}$ for $\mu=x,y,z$ at $H=0.75$, $K=0.3$, $I^z=1.2$, $I^{xy}=0.05$, and $I^v=0$ for (a,b) $\phi=0$ and (c,d) $\phi=\pi/4$.  
}
\end{center}
\end{figure}

\begin{figure}[htb!]
\begin{center}
\includegraphics[width=1.0 \hsize ]{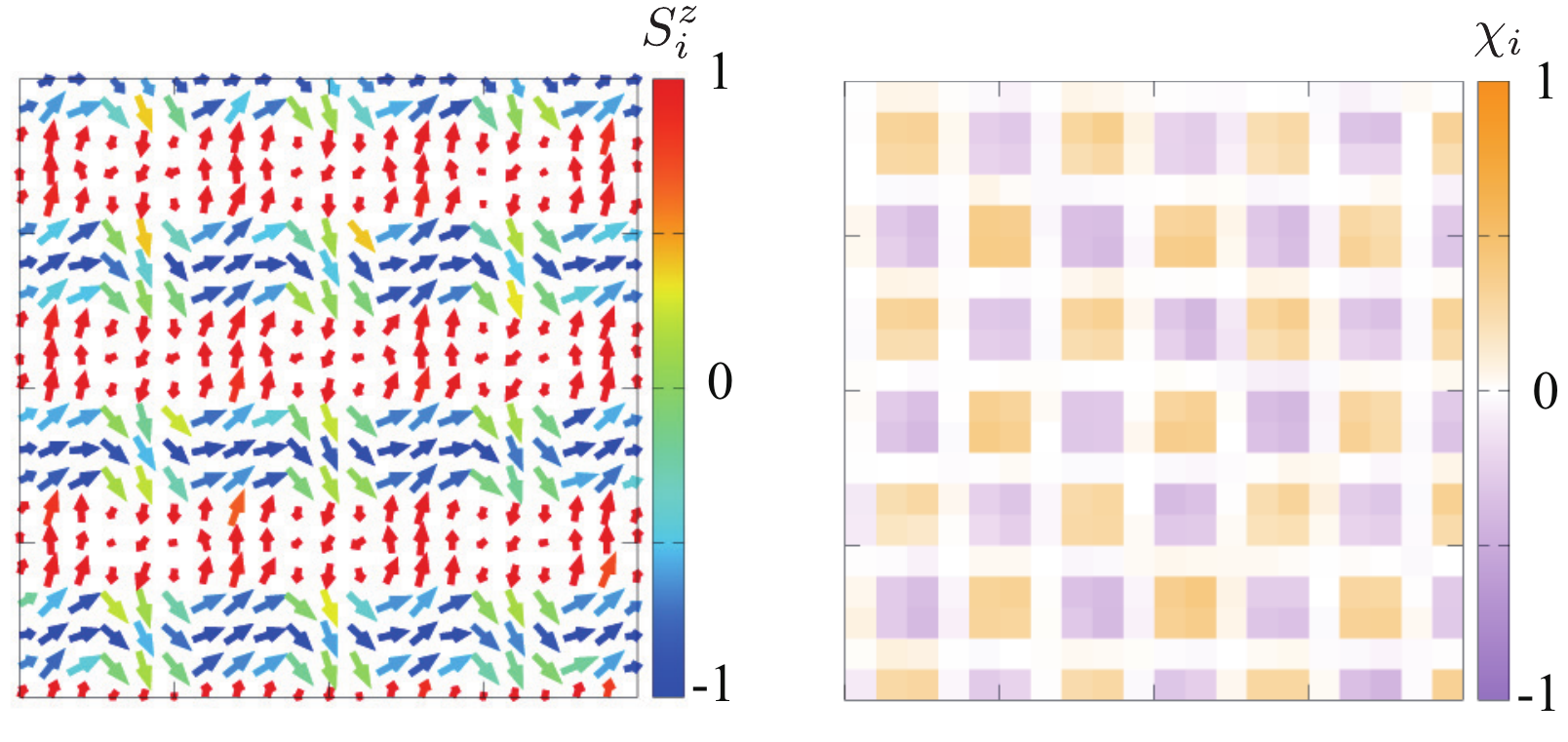} 
\caption{
\label{fig:spin_rot}
Left: Snapshots of the spin configurations at $\theta=45^{\circ}$ and $\phi=0$. 
The direction and the color of the arrows represent the $xy$ and $z$ components of the spin moment, respectively. 
Right: Snapshots of the scalar chirality configuration $\chi_i$ calculated from the left panel. 
}
\end{center}
\end{figure}

\begin{figure}[t!]
\begin{center}
\includegraphics[width=1.0 \hsize ]{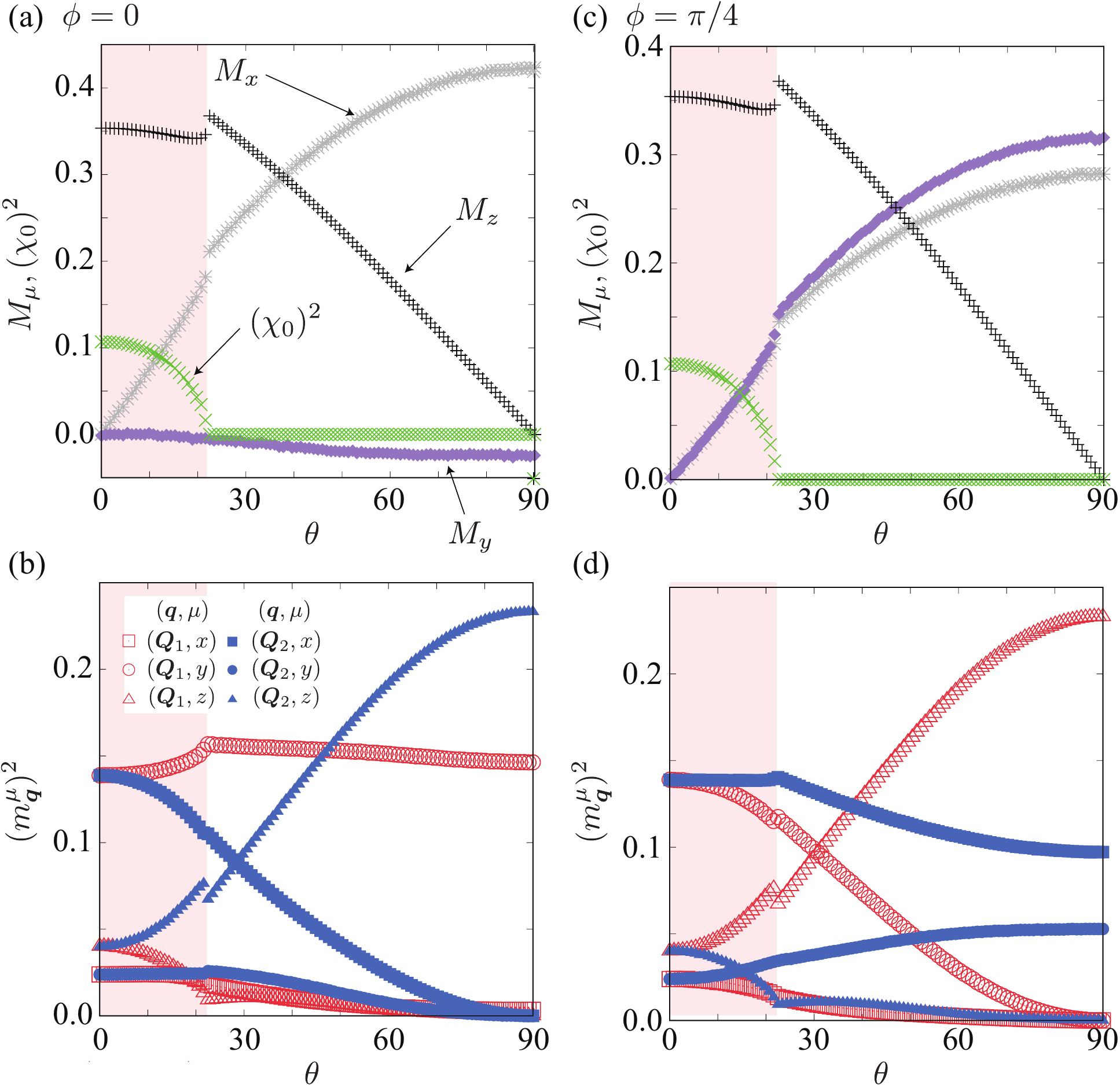} 
\caption{
\label{fig:Mq_rotH_2}
$\theta$ dependences of (a,c) $M_\mu$, $(\chi_0)^2$, (b,d) $m^\mu_{\bm{Q}_1}$, and $m^\mu_{\bm{Q}_2}$ for $\mu=x,y,z$ at $H=0.75$, $K=0.3$, $I^z=1.2$, $I^{xy}=0.05$, and $I^v=0.05$ for (a,b) $\phi=0$ and (c,d) $\phi=\pi/4$.  
}
\end{center}
\end{figure}

Finally, let us discuss the stability of the square SkX in a magnetic field rotation, i.e., $\theta \neq 0$. 
We consider two cases in a rotated field, $\phi=0$ and $\phi=\pi/4$, while changing $\theta$.
We first discuss the situation with $I^{xy}=0.05$ and $I^v=0$.
The other model parameters are taken at $H=0.75$, $K=0.3$, and $I^z=1.2$, where the square SkX is stabilized for $\theta=0$ as discussed in Sec.~\ref{sec:Skyrmion crystal in an out-of-plane field}. 

Figures~\ref{fig:Mq_rotH_1}(a) and \ref{fig:Mq_rotH_1}(b) show the $\theta$ dependences of the magnetization $M_{\mu}$ and the scalar chirality $(\chi_0)^2$ and the $\bm{Q}_\eta$ component of the magnetic moment $m^\mu_{\bm{Q}_\eta}$ for $\mu=x,y,z$, respectively. 
When the magnetic field is tilted from $\theta=0$ to the $x$ direction ($\phi=0$), $(\chi_0)^2$ gradually decreases in Fig.~\ref{fig:Mq_rotH_1}(a). 
In addition, $m^\mu_{\bm{Q}_\eta}$ perpendicular (parallel) to the $x$ axis increases (decreases) to gain the Zeeman energy, as shown in Fig.~\ref{fig:Mq_rotH_1}(b). 
Thus, the SkX is deformed in a rectangle way. 
While increasing $\theta$, the SkX is replaced by the other double-$Q$ state at $\theta \simeq 24^{\circ}$. 
Although this double-$Q$ state seems to have similar $m^\mu_{\bm{Q}_\eta}$ to the SkX, but this state does not have $(\chi_0)^2$; the local scalar chirality is distributed in a checkerboard way, where the real-space spin and chirality configurations are shown in the left and right panels of Fig.~\ref{fig:spin_rot}, respectively. 
With a further increase of $\theta$, the amplitude of $m^\mu_{\bm{Q}_1}$ and $m^\mu_{\bm{Q}_2}$ becomes different for $\theta \gtrsim 63^{\circ}$. 
This anisotropic double-$Q$ state exhibits a nonzero net magnetization along the $y$ direction, as shown in Fig.~\ref{fig:Mq_rotH_1}(a), which is owing to the inequivalence between the ${\bm{Q}_1}$ and $\bm{Q}_2$ components of spins under the inplane magnetic field.

Meanwhile, when the magnetic field is tilted from the $z$ direction to the $[110]$ direction ($\phi=\pi/4$), there are no isotropic double-$Q$ states while varying $\theta$, as shown in Figs.~\ref{fig:Mq_rotH_1}(c) and \ref{fig:Mq_rotH_1}(d). 
The intensities of $m^\mu_{\bm{Q}_1}$ and $m^\mu_{\bm{Q}_2}$ become different for infinitesimally small $\theta$. 
It is noted that the $x$ and $y$ components of $M_{\mu}$ and $m^\mu_{\bm{Q}_\eta}$ show the same behavior, as we only consider the anisotropic magnetic interactions $I^{xy}$. 

Next, we discuss the case for $I^{xy}=I^{v}=0.05$, where the results for $\phi=0$ and $\phi=\pi/4$ are shown in Figs.~\ref{fig:Mq_rotH_2}(a), \ref{fig:Mq_rotH_2}(b) and Figs.~\ref{fig:Mq_rotH_2}(c), \ref{fig:Mq_rotH_2}(d), respectively. 
In both cases for $\phi=0$ and $\phi=\pi/4$, a single phase transition occurs from the SkX to the anisotropic double-$Q$ state, which is similar to the result in Fig.~\ref{fig:Mq_rotH_1}(d). 
In other words, the isotropic double-$Q$ state shown in Fig.~\ref{fig:spin_rot} does not appear for $I^{xy}=I^{v}=0.05$. 
As both [100] and [110] directions are not high-symmetry lines in the presence of both $I^{xy}$ and $I^{v}$, the inplane magnetizations, $M_x$ and $M_y$, are different from each other. 

\section{Summary}
\label{sec:Summary}

To summarize, we have investigated the effect of magnetic anisotropic interactions that originate from the lacking of the mirror symmetry in the centrosymmetric tetragonal crystal systems on the formation of the square SkXs. 
Through the analyses by the simulated annealing for the effective spin model with the bilinear and biquadratic interactions in momentum space on the two-dimensional square lattice, we mainly discussed two important features in the present system: 
One is that the anisotropic interaction in the form of $I^{xy}(S^x_{\bm{Q}_\eta}S^y_{-\bm{Q}_\eta}+S^y_{\bm{Q}_\eta}S^x_{-\bm{Q}_\eta})$ can be a microscopic origin of the square SkX in centrosymmetric itinerant magnets. 
The other is that the helicity of the SkXs is fixed by two types of anisotropic interactions, $I^{xy}$ and $I^v$. 
We have shown that different types of active odd-parity multipoles appear for the different helicity, where odd-parity magnetic and magnetic toroidal multipoles are related to the linear magnetoelectric effect and odd-parity electric and electric toroidal multipoles are related to the antisymmetric spin-split band structure and the Edelstein effect. 
We have also discussed the stability and the related phase transitions of the SkXs in the magnetic field rotation. 
We found the isotropic double-$Q$ state without the net scalar chirality in the [101] magnetic field when $I^v=0$.

The helicity locking of the magnetic skyrmion in centrosymmetric magnets with the anisotropic interactions can extend the scope of the application to the skyrmion-based racetrack memories.
Our results regarding the helicity locking in the presence of magnetic anisotropy can be applied to the isolated skyrmion, which is important from the viewpoint of practical applications~\cite{zhang2015magnetic,zhang2020skyrmion}.
One of the challenges for the application is to realize the situation where the skyrmion moves in a parallel direction to an external electric current without moving to the perpendicular direction like the skyrmion Hall effect.
Although such a situation has been usually discussed in antiferromagnetic skyrmions~\cite{PhysRevLett.116.147203,zhang2016antiferromagnetic}, our result indicates that the lattice systems without vertical mirror symmetry might be an alternative system by avoiding the skyrmion Hall effect based on the helicity degree of freedom.
In fact, such an attempt of controlling the skyrmion transport by using the helicity degree of freedom has been studied in noncentrosymmetric systems~\cite{Jin_APL10.1063/1.5095686}, which can be extended to the centrosymmetric systems in the present model. 
The candidate materials are GdRu$_2$Si$_2$~\cite{khanh2020nanometric,Yasui2020imaging,khanh2022zoology}, EuAl$_4$~\cite{Shang_PhysRevB.103.L020405,kaneko2021charge,Zhu2022}, EuGa$_4$~\cite{zhang2022giant,Zhu2022}, and EuGa$_2$Al$_2$~\cite{moya2021incommensurate}, where the square SkX was observed and suggested in experiments. 
Since the crystal structures in these compounds belong to the $D_{\rm 4h}$ point group, the symmetry lowering by external stimuli, such as chemical doping, is required~\cite{Tokura_doi:10.1021/acs.chemrev.0c00297}. 

The present result provides a further possibility of the SkXs induced by the anisotropic interactions that arise from the mirror symmetry breaking. 
The similar SkX and its helicity locking can be expected to occur in the hexagonal point group $C_{\rm 6h}$. 
Furthermore, although we have shown that the different helicity in the SkXs leads to different physical phenomena, a similar argument would hold for the different multiple-$Q$ states with the helicity degree of freedom, which are stabilized by the anisotropic interactions, such as the hedgehog lattice~\cite{kato2021spin} and meron-antimeron crystal~\cite{Hayami_PhysRevB.104.094425}.

\begin{acknowledgments}
This research was supported by JSPS KAKENHI Grants Numbers JP19K03752, JP19H01834, JP21H01037, and by JST PRESTO (JPMJPR20L8).
R.Y. was supported by Forefront Physics and Mathematics Program to Drive Transformation (FoPM).
Parts of the numerical calculations were performed in the supercomputing systems in ISSP, the University of Tokyo.
\end{acknowledgments}

\bibliographystyle{apsrev}
\bibliography{ref}

\begin{thebibliography}{149}
\bibitem[{com({\natexlab{a}})}]{comment_Mu}
\bibinfo{note}{Although the $3z^2-r^2$ type of the magnetic quadrupole $M_u$
  also appears when $M_0$ is active, we do not explicitly show it because we
  consider the two-dimensional response tensor.}
\bibitem[{com({\natexlab{b}})}]{comment_Gu}
\bibinfo{note}{Although the $3z^2-r^2$ type of the electric toroidal quadrupole
  $G_u$ also appears when $G_0$ is active, we do not explicitly show because we
  consider the two-dimensional band structure and the response tensor.}
\end{thebibliography}

\end{document}